\documentclass[nofootinbib,aps,prd,twocolumn,showpacs,preprintnumbers,amsmath,amssymb]{revtex4}
\usepackage{umlaut}
\usepackage{epsfig}
\usepackage{longtable}
  \newcommand{\tr}{{\rm Tr}}
\def\lsim{\raise0.3ex\hbox{$<$\kern-0.75em\raise-1.1ex\hbox{$\sim$}}}
\def\gsim{\raise0.3ex\hbox{$>$\kern-0.75em\raise-1.1ex\hbox{$\sim$}}}

\def\eg{{\sl e.\ g.\/}}
\def\etc{{\sl etc.\/}}

\def\ie{{\sl i.e.\/}}
\def\beq{\begin{equation}}
\def\eeq{\end{equation}}
\def\beqa{\begin{eqnarray}}
\def\eeqa{\end{eqnarray}}
\def\re{{\rm Re\/}}

\def\Z{{Z}}

\begin{document}
\title{Renormalized Polyakov loops in many representations}

\author{Sourendu Gupta}\email{sgupta@theory.tifr.res.in}
\affiliation{Department of Theoretical Physics, Tata Institute for Fundamental Research,\\
 Homi Bhabha Road, Mumbai 400 005, India}

\author{Kay H\"{u}bner}\email{huebner@bnl.gov}
\affiliation{Physics Department, Brookhaven Natl. Laboratory, Upton, New York 11973, USA}

\author{Olaf Kaczmarek}\email{okacz@physik.uni-bielefeld.de} 
\affiliation{Fakult\"{a}t f\"{u}r Physik, Universit\"{a}t Bielefeld, D-33615 Bielefeld, Germany}

\date{\today} \preprint{BI-TP 2007/30, BNL-NT-07/45 and TIFR/TH/07-30}


\begin{abstract}
We present a renormalization procedure for Polyakov loops which explicitly
implements the fact that the renormalization constant depends only on the
ultraviolet cutoff. Using this we study the renormalized Polyakov loops in
all representations upto the {\bf 27} of the gauge group SU(3).
We find good evidence for Casimir scaling of the Polyakov loops and for
approximate large-$N$ factorization.
By studying many loops together, we are able to show
that there is a matrix model with a single coupling which can describe
the high temperature phase of QCD, although it is hard to construct
explicitly. 
We present the first results for the non-vanishing renormalized octet loop
in the
thermodynamic limit below the SU(3) phase transition, 
and estimate the associated string breaking distance
and the gluelump binding energy.
By studying the connection of the direct renormalization procedure with
a generalization of an earlier suggestion which goes by the name of the
$Q\overline Q$ renormalization procedure, we find that they are functionally
equivalent.
\end{abstract}

\maketitle

\section{Introduction}\label{Sec1}
The proof of confinement in QCD is now literally a million dollar question
\cite{clay}. There are many ideas about the direction in which such
a proof lies. No matter what these ideas are, once they are properly
formulated, they are always open to test by lattice techniques. One of
the long-lasting ideas has been to examine a toy model of QCD for large
number of colors, $N$ \cite{tHooft:1973jz}.  Because of the enhanced symmetry,
many quantities become amenable to study in this limit. Computations of
corrections upto sub-leading order, $1/N$, have been made for quantities
such as hadron masses and pion-nucleon scattering, with intriguing
results. Recently, by adding supersymmetries to large-$N$ QCD, very simple toy
models have been constructed which are amenable to analytical treatment
using the AdS/CFT correspondence \cite{Policastro:2001yc}.  Much excitement has
been generated by the plethora of predictions of such toy models, and
there have been exciting speculation about their applicability to QCD.\\
At large $N$ the dynamics of quarks is secondary to that of the gluons,
being suppressed by power corrections in $N$. Hence lattice tests of
these ideas have been made in pure gauge, or quenched, QCD. There have
been investigations of the string tension and its scaling with $N$
\cite{Bringoltz:2006zg}, the nature of the phase transition with changing $N$
\cite{Lucini:2003zr}, and tests of approximate scale-invariance of finite
temperature QCD \cite{Gavai:2004se}.  In this paper we investigate certain
ideas about Polyakov loops and their behavior at finite temperature
which have developed in recent years based on large $N$ and the AdS/CFT
correspondence. We investigate matrix models which could be expected to
describe the high temperature phase of pure gauge QCD. Our results put
very strong constraints on the kinds of matrix models which may provide
a description of pure gauge QCD.\\
The Polyakov loop is closely connected with
confinement since it is the order parameter for
the transition from a confined to a
deconfined medium.
Various models based on Polyakov loops have been proposed to
describe the transition to a quark gluon plasma phase and its properties at
zero as well as non-zero baryon density
\cite{Pisarski:2006hz,Dumitru:2003hp,Dumitru:2004gd,
    Dumitru:2005ng,Dumitru:2003cf,Megias:2005ve,Megias:2004hj,Ratti:2005jh,
    Meisinger:2003id,Fukushima:2003fm,Diakonov:2004kc,Schaefer:2007pw,Blaschke:2007np,Ghosh:2007,Kashiwa:2007hw}.
Furthermore the connection of SU(3) theory to the
large $N$ limit, in the mean-field approximation, was discussed in
\cite{Dumitru:2003hp,Dumitru:2004gd}.
For a test of the reliability and comparison of these Polyakov loop models to
pure gauge theory and QCD with dynamical quarks, a detailed knowledge of the
behavior of the Polyakov loop in the fundamental and higher
representations in those theories is needed.\\
The Polyakov loop needs to be renormalized, since it has divergent
contributions from the ultraviolet. In Section \ref{sec:renormpl}
we present a renormalization procedure which explicitly incorporates
properties expected of a good scheme. This direct renormalization
technique is naturally applicable to Polyakov loop expectation
values in all representations of the color group. The multiplicative
renormalizations in different representations are closely connected if
the Polyakov loops satisfy a property called Casimir scaling.  We present
tests of Casimir scaling in Section \ref{sec:casimir}. This leads on,
in Section \ref{sec:large_nc}, to a test of large-$N$ factorization at
$N=3$. We find good evidence for both in a temperature range not too
close to $T_c$.\\
Next, in Section \ref{matrixmodel}, we examine whether the renormalized
Polyakov loops are described in an effective matrix
model. By examining the renormalization scheme dependence of these
quantities, we find that a single parameter variation of matrix models
describes the temperature dependence of the Polyakov loops in various
representations. We also show that the matrix model is unlikely to consist
of a small set of terms, and therefore hard to construct explicitly from
the phenomenology of Polyakov loops.\\
In Section \ref{sec_strinbreaking} we consider the adjoint Polyakov
loop correlations below $T_c$. We report the first measurement of the
renormalized adjoint Polyakov loops in the thermodynamic limit at finite
temperature in the confined phase of QCD. We find that aspects of the
adjoint Polyakov loop correlations can be summarized in the physics of
gluelumps, \ie, colorless states made of static adjoint sources and glue.\\
In the appendices we consider a renormalization procedure, the $Q\overline
Q$ procedure, earlier suggested in \cite{Kaczmarek:2002mc}. We
extend it to the renormalization of Polyakov loops in arbitrary
representations, consider the relation between the direct and $Q\overline
Q$ renormalization procedures, and examine ``color averaged'' Polyakov
loop correlators in various representations.

\section{Details of the calculations}
\label{sec:sim}
We have performed simulations for temperatures up to
$24~T_c$ on
$N_\sigma^3\times N_\tau$ lattices with $N_\tau=4,6,8$ and
$N_\sigma$ up to 32 in SU(3) pure gauge
theory with the tree level Symanzik-improved gauge action
\cite{Weisz:1982zw,Weisz:1983bn}. 
We used a pseudo heatbath algorithm with FHKP updating in the SU(2)
subgroups. Each heatbath update was followed by four overrelaxation steps. 
The statistics varies from 1000 to 10000 of such sweeps after suitable
thermalization.
The physical scale has been set using the zero temperature string tension,
$\sigma$, \cite{Beinlich:1997ia}
and a determination of the critical coupling for the deconfinement transition
from \cite{Beinlich:1996xg}. 
We have calculated the Polyakov loops in all representations up to $D=27$ using
the operators defined in (\ref{pol1})-(\ref{pol7}).
The errors on the observables were determined with the Jackknife method.\\
Furthermore, we have reanalyzed
configurations generated using two flavors of staggered quarks with mass $m/T=0.4$ on
a $16^3\times 4$ lattice at several temperatures above and below the
transition temperature \cite{Allton:2002zi,Allton:2003vx}. At each temperature we have used
statistics of several thousands to
calculate Polyakov loops up to $D=27$.

\section{Renormalization of Polyakov loops}
\label{sec:renormpl}
We define the thermal Wilson line, $P(\vec x)$ at spatial position $\vec x$ as
\begin{eqnarray}
P(\vec x) \equiv \prod_{i=0}^{N_\tau-1} U_{(\vec x,i),0},
\label{wilsonline}
\end{eqnarray}
where $U_{(\vec x,i),0}$ is the gauge link matrix in the time direction
at the point $\vec x$ and Euclidean time $i$. $U$ is a $3\times3$ matrix
which belongs to $SU(3)$. 
We define the local Polyakov loop as 
the trace of $P(\vec x)$, 
\begin{eqnarray}
L(\vec x) \equiv \tr P(\vec x),
\end{eqnarray}
where the trace is normalized to one.
We denote the expectation value of the Polyakov loop
in the fundamental representation of SU(3) by
\begin{eqnarray}
L_3 \equiv \Big\langle \frac{1}{V}\sum_{\vec x} L(\vec x) \Big\rangle.
\end{eqnarray}
Polyakov loops in different representations, $L_D$, are defined in Appendix
\ref{su3}.
The subscript $D$ indicates the dimension of the
color (irreducible) representation of the Polyakov loop, \eg, $L_3$ for fundamental or $L_8$
for adjoint.
Expectation values of Polyakov loops are ultraviolet divergent.
We will use the superscripts $b$ or $r$ for bare 
and renormalized Polyakov loop respectively.
\subsection{Basic properties of renormalization}\label{intro_ren}
It was pointed out by Polyakov \cite{Polyakov:1980ca} that for smooth
loops, ultraviolet divergences can be absorbed in the charge
renormalization of gauge fields:
\begin{eqnarray}
L_D^{r}(T) = \left(\Z_D(g^2)\right)^{\ell(C)} L_D^{b}(g^2),
\label{pol_rel}
\end{eqnarray}
where $\ell(C)$ is the length of the contour and the coupling, $g^2=6/\beta$, on the right is the
bare coupling. The quantity on the left is properly renormalized and depends on the
renormalized coupling or, through this running coupling, on the temperature. 
Cusps and self intersections of loops give rise to
logarithmic divergences which depend, \eg, on the angle
of the cusps \cite{Polyakov:1980ca,Dotsenko:1979wb,Korchemsky:1987wg}.
Spatial averages of operators such as $\tr [L(\vec x)]^n$, which wind $n$
times around the lattice, also need separate renormalization. Similarly,
composite operators such as powers of Polyakov loops,
including Polyakov loop susceptibilities, also
require independent renormalization.\\
We have written eq.\ (\ref{pol_rel}) for
an arbitrary representation $D$. The renormalization constants in different
representations, $\Z_D$, can be related to each other if
both the bare and the renormalized loops satisfy the relation
\begin{eqnarray}
L_D^{1/{C_2(D)}} = L_{D'}^{1/{C_2(D')}},
\label{casimirscaling}\end{eqnarray}
called Casimir scaling.
Here $C_2(D)=\tr\sum_a\lambda^a\lambda^a$ is the quadratic Casimir operator
in the representation $D$.
When Casimir scaling holds, the quantities
\beq
  Z_D(g^2) = \left[\Z_3(g^2)\right]^{1/d_D},
\label{renormcs}\eeq
where $d_D=C_2(D)/C_2(3)$, are all equal.\\
In keeping with the general form in eq.\ (\ref{pol_rel}) we can write
\beq
L_D^{r}(T) = \left(Z_D(g^2)\right)^{d_DN_\tau} L_D^{b}(g^2,N_\tau)
\label{pol_rel2}
\eeq
where the renormalization constants $Z_D(g^2)$ should only depend on the
bare coupling. Such a multiplicative renormalization is expected to compensate
entirely for the dependence of the bare loop on the cutoff, \ie, the bare
coupling, so that the renormalized loop on the left is a function only of the
temperature. Furthermore, if Casimir scaling is found to hold, then all the $Z_D$
collapse to a single function $Z_3(g^2)$, i.e. loops in all
representations can be simultaneously renormalized. Note that there is one
remaining ambiguity: the function $Z_3(g^2)$ can be multiplied by a single
coupling independent constant without affecting the renormalization. Thus,
a one parameter family of renormalization schemes for Polyakov loops is
defined by eq.\ (\ref{pol_rel2}).

\subsection{Direct renormalization of the Polyakov loop}\label{sec_direct}
\begin{figure}[thb]
\begin{center}
\scalebox{0.65}{\includegraphics{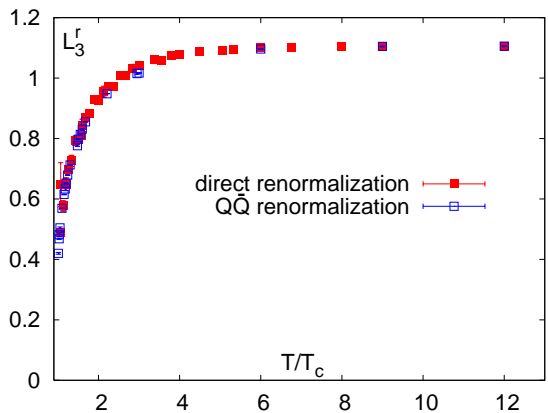}}
\end{center}
\caption{
Comparison of the renormalized Polyakov loop in the fundamental representation,
$L_{3}^{r}(T)$, obtained with the two different renormalization procedures. 
}
\label{fig:renpol3}
\end{figure}
\begin{figure}[thb]
\begin{center}
\scalebox{0.65}{\includegraphics{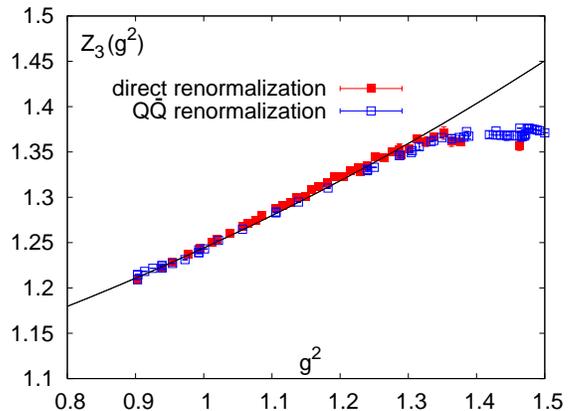}}
\end{center}
\caption{
Comparison of the renormalization constants for fundamental loops,
$Z_{3}^R(g^2)$, for the two different renormalization procedures. $g^2$
denotes the bare coupling and the solid line is the same as in fig.~\ref{fig:renconst48}.
}
\label{fig:renconst3}
\end{figure}
In this subsection we present a complete renormalization procedure which
implements eq.\ (\ref{pol_rel2}).  We call this the direct renormalization
prescription. A similar procedure was discussed
earlier in \cite{Creutz:1980hb}.  Denote by $L_D(g^2,N_\tau)$ the Polyakov
loop expectation value obtained after taking the thermodynamic limit
at a temperature $T=1/a(g^2)N_\tau$, where $a$ is the lattice spacing
at a bare coupling $g^2$ and $N_\tau$ is the temporal extent of the
lattice. We will describe the procedure for a fundamental loop first.\\
First choose the value of $L_3^{r}(T_{ref})$, at a reference temperature
$T_{ref}$. It is clear from eq.\ (\ref{pol_rel2}) that this choice is
exactly equivalent to fixing the renormalization scheme. It is convenient,
but not necessary, to take $T_{ref}$ to be the maximum temperature in
the study: in our case $T_{ref}=12 T_c$. We discuss our choice of scheme
in Appendix \ref{sec:scale}.  This sets the first step of the iterative
procedure starting at the initial temperature $T_1=T_{ref}$.\\
Next we need measurements at (at least) two different temporal
extents, $N_\tau^\alpha$ and $N_\tau^\beta$, say, with
$N_\tau^\alpha>N_\tau^\beta$, both at the temperature
$T_i$ (we begin with $i=1$ and set up an iteration). Therefore these
measurements correspond to two different lattice cutoffs
$a(g^2_{i,\alpha}) N^\alpha_\tau=a(g^2_{i,\beta}) N^\beta_\tau$ with
$a_{i,\alpha}<a_{i,\beta}$, where the subscripts are self-explanatory.
We obtain two different renormalization constants,
\beqa
   \left(Z_3(g^2_{i,\alpha})\right)^{N^\alpha_{\tau}}
      L^{b}_3(g^2_{i,\alpha},N^\alpha_{\tau})
   &=& L^{r}_3(T_i),\label{pren1} \\
   \left(Z_3(g^2_{i,\beta})\right)^{N^\beta_{\tau}}
      L^{b}_3(g^2_{i,\beta},N^\beta_{\tau})
   &=& L^{r}_3(T_i).\label{pren2}
\eeqa
The third step is to advance the iteration. We do this by making a
measurement of $L^{b}_3(g^2_{i,\beta},N^\alpha_\tau)$ on the lattice
with temporal extent $N^\alpha_\tau$ at a temperature $T_{i+1}=1/a_{i,\beta}
N^\alpha_\tau=(N^\beta _\tau/N^\alpha_\tau)T_i$.  Since the renormalization
constant is already known from eq.\ (\ref{pren2}), one obtains
the value of $L_3^{r}(T_{i+1})$
\beq
   L^{r}_3(T_{i+1}) = \left(Z_3(g^2_{i,\beta})\right)^{N^\alpha_{\tau}}
      L^{b}_3(g^2_{i,\beta},N^\alpha_\tau).
\eeq
Since we have the value of the renormalized loop at a new temperature,
we can now iterate the procedure from the second step on. The iteration
gives the renormalized loops and the renormalization constants at a
decreasing series of temperatures.\\
Four points about the prescription are worth noting explicitly.
First, the procedure extends without change to any representation
$D$. The test of Casimir scaling would be to assume that the bare
loops in different representations at $T_{ref}$ are related by eq.\
(\ref{casimirscaling}), and then check whether the renormalized loops
at all $T$ are related in the same way. We discuss this further in
Section \ref{sec:casimir}. Second, in the confined phase of
the pure gauge theory the bare Polyakov loop, in any representation
with non-vanishing triality, vanishes in the thermodynamic limit;
as a result the direct renormalization procedure can only be used
above $T_c$ for such representations.  Third, a reverse iteration can
always be performed by choosing $T_{i+1}=1/a_{i,\alpha} N^\alpha_\tau =
(N^\alpha_\tau/N^\beta_\tau) T_i> T_i$.  Finally, although we discussed
the procedure for two values of $N_\tau$, it can be easily extended to
a larger number of values for the temporal extent.\\
The renormalized Polyakov loop in the fundamental representation
obtained by the direct procedure described here is shown in Figure
\ref{fig:renpol3}.  Also shown, for comparison, are the results
obtained from a completely different renormalization procedure
\cite{Kaczmarek:2002mc} based on a matching of the short distance behavior
of heavy quark-antiquark free energies to the zero temperature potential
(labeled as $Q\bar Q$-renormalization). Both these procedures allow
a one parameter family of renormalization schemes, and the schemes
have been chosen so that the value of $L_3^{(r)}(T_{ref})$ match.
Figure \ref{fig:renconst3} shows the results for the renormalization
constant.  These figures indicate the functional equivalence of the two
renormalization procedures.

\subsection{Fundamental Polyakov loops}
\label{sec_fundamental}
\begin{figure}[t]
\begin{center}
\epsfig{file=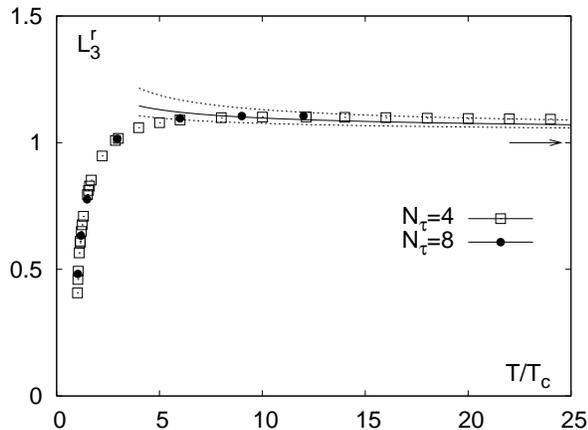,width=8.65cm}
\end{center}
\caption{ The renormalized fundamental Polyakov loop in SU(3) pure gauge theory
  for two values of the
  temporal lattice extend $N_\tau$. The lines show the perturbative result
  (\ref{Gava}) and are explained in the text. The
  arrow represents the asymptotic high
  temperature limit, $L_3^{r}=1$.}
\label{fig:Lren_PGpaper2}
\end{figure}
\begin{figure}[t]
\begin{center}
\epsfig{file=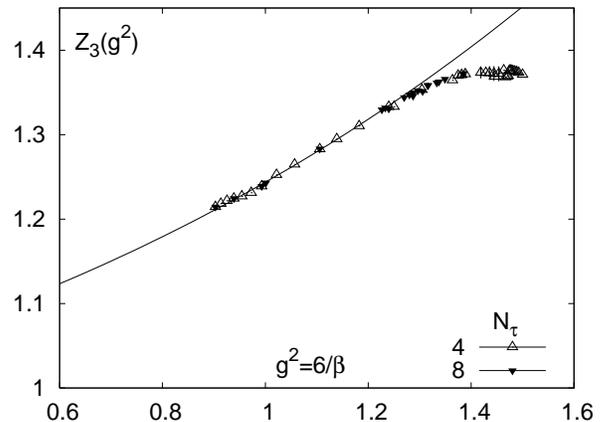,width=8.65cm}
\end{center}
\caption{The renormalization constants $Z_3(g^2)$ as a function of the bare
  coupling $g^2=6/\beta$ calculated on lattices of size $32^3\times N_\tau$ with
  $N_\tau=4$ and 8. The line comes from a fit to (\ref{z_pert}) as explained in
  the text.}
\label{fig:renconst48}
\end{figure}
\begin{table}[thbp]
\centering
\setlength{\tabcolsep}{0.5pc}
\begin{tabular}{|c|l|l||c|l|l|}
\hline
$N_\tau$ & $T/T_c$ & $L_3^{r}$ & $N_\tau$ & $T/T_c$ & $L_3^{r}$  \\
\hline
4 & 1.012  &  0.4070(11)      &   4 & 6.001  &  1.0897(4)      \\ 
4 & 1.031  &  0.4600(7)       &   4 & 8.002  &  1.0986(5)      \\ 
4 & 1.049  &  0.4927(22)      &   4 & 10.00  &  1.1011(7)      \\ 
4 & 1.099  &  0.5649(14)      &   4 & 12.13  &  1.1014(6)      \\ 
4 & 1.144  &  0.6049(3)       &   4 & 14.00  &  1.1000(3)      \\ 
4 & 1.151  &  0.6114(16)      &   4 & 16.00  &  1.0988(6)      \\ 
4 & 1.200  &  0.6494(12)      &   4 & 18.01  &  1.0966(5)      \\ 
4 & 1.241  &  0.6759(15)      &   4 & 20.00  &  1.0954(8)      \\ 
4 & 1.301  &  0.7095(13)      &   4 & 22.00  &  1.0939(10)     \\ 
4 & 1.499  &  0.7953(13)      &   4 & 24.00  &  1.0924(12)     \\ 
4 & 1.549  &  0.8115(8)       &     &        &                 \\
4 & 1.600  &  0.8288(9)       &   8 &   1.03  &  0.4818(99)    \\ 
4 & 1.684  &  0.8523(2)       &   8 &   1.18  &  0.6330(125)   \\ 
4 & 2.214  &  0.9475(3)       &   8 &   1.48  &  0.7763(116)   \\ 
4 & 2.858  &  1.0087(1)       &   8 &   2.95  &  1.0149(68)    \\ 
4 & 2.999  &  1.0169(1)       &   8 &   6.00  &  1.0961(33)    \\ 
4 & 3.987  &  1.0591(2)       &   8 &   9.00  &  1.1049(27)    \\ 
4 & 5.001  &  1.0791(2)       &   8 &   12.00 &  1.1060(26)    \\ 
\hline
\end{tabular}
\caption{The renormalized fundamental Polyakov loop $L_3^{r}(T)$ obtained on
  lattices of size $32^3\times N_\tau$ with $N_\tau=4$ and $8$. $T/T_c$ denotes
the temperature in units of the critical temperature.}
\smallskip
\label{table_fpol}
\end{table}

\begin{table}[thbp]
\centering
\setlength{\tabcolsep}{0.45pc}
\begin{tabular}{|c|l|l||c|l|l|}
\hline
$N_\tau$ & $g^2$ & $Z_3$ & $N_\tau$ & $g^2$ & $Z_3$  \\
\hline
4 & 0.90294 &  1.2144(2)      &         4 & 1.47783 &  1.3759(1)        \\  
4 & 0.91348 &  1.2183(3)      &         4 & 1.48148 &  1.3754(1)        \\  
4 & 0.92531 &  1.2217(2)      &         4 & 1.48515 &  1.3748(1)        \\  
4 & 0.93869 &  1.2245(3)      &         4 & 1.48883 &  1.3742(1)        \\  
4 & 0.95426 &  1.2270(2)      &         4 & 1.49254 &  1.3733(1)        \\  
4 & 0.97248 &  1.2312(1)      &         4 & 1.50000 &  1.3711(1)        \\  
4 & 0.99282 &  1.2389(2)      &           &         &                   \\
4 & 1.02157 &  1.2525(1)      &         8 &  0.90294  &  1.2145(2)      \\   
4 & 1.05684 &  1.2648(1)      &         8 &  0.93882  &  1.2246(2)      \\   
4 & 1.10577 &  1.2829(2)      &         8 &  0.99282  &  1.2391(2)      \\   
4 & 1.13889 &  1.2948(1)      &         8 &  1.00117  &  1.2431(1)      \\   
4 & 1.18227 &  1.3102(2)      &         8 &  1.10577  &  1.2833(2)     \\    
4 & 1.23993 &  1.3330(13)     &         8 &  1.22647  &  1.3297(21)    \\    
4 & 1.25000 &  1.3331(2)      &         8 &  1.23302  &  1.3319(23)    \\    
4 & 1.30435 &  1.3531(9)      &         8 &  1.23985  &  1.3307(2)     \\    
4 & 1.36364 &  1.3646(1)      &         8 &  1.26995  &  1.3441(36)    \\    
4 & 1.37457 &  1.3700(39)     &         8 &  1.28003  &  1.3473(33)    \\    
4 & 1.38857 &  1.3717(47)     &         8 &  1.28703  &  1.3459(1)     \\    
4 & 1.41878 &  1.3734(68)     &         8 &  1.28742  &  1.3496(31)    \\    
4 & 1.42857 &  1.3731(2)      &         8 &  1.29618  &  1.3523(29)    \\    
4 & 1.43575 &  1.3730(80)     &         8 &  1.30574  &  1.3514(1)     \\    
4 & 1.44439 &  1.3724(87)     &         8 &  1.31579  &  1.3585(26)    \\    
4 & 1.45384 &  1.3721(88)     &         8 &  1.31602  &  1.3583(28)    \\    
4 & 1.46699 &  1.3681(8)      &         8 &  1.33333  &  1.3614(2)     \\    
4 & 1.47059 &  1.3690(2)      &         8 &  1.33743  &  1.3628(3)     \\    
4 & 1.47420 &  1.3757(2)      &         8 &  1.34916  &  1.3659(3)     \\    
4 & 1.47601 &  1.3761(1)      &         8 &  1.38530  &  1.3724(2)     \\    
\hline
\end{tabular}
\caption{The renormalization constants for the fundamental Polyakov loop,
  $Z_3(g^2)$ obtained on
  lattices of size $32^3\times N_\tau$ with $N_\tau=4$ and $8$. $g^2=6/\beta$
  denotes the bare coupling.}
\smallskip
\label{table_fz}
\end{table}

We have extended previous measurements of the fundamental Polyakov loop
\cite{Kaczmarek:2002mc} to temperatures as high as $24 T_c$.  The results
for $L_3^{r}(T)$ are shown in Figure \ref{fig:Lren_PGpaper2} and listed
in Table \ref{table_fpol}. The corresponding renormalization constants are
plotted in Figure \ref{fig:renconst48} and listed in Table \ref{table_fz}.
The direct renormalization procedure for the fundamental Polyakov loop
stops at $g^2$ corresponding to $T_c$ on the lattice with the smallest
$N_\tau$. Since the $Q\overline Q$ procedure gives identical results upto
this point, and can be continued to larger $g^2$, the tables contain results
obtained using this procedure.\\
The Polyakov loop for SU(N) pure gauge theory in HTL
perturbation theory \cite{Gava:1981qd} is
\beqa
\nonumber
L_D &=& 1+2\pi^2 C_2(D) \Big\{\left(\frac{2}{3}N\right)^{\frac{1}{2}}\left(\frac{g^2}{8\pi^2}\right)^{\frac{3}{2}}\\
\nonumber
&&\qquad\qquad+N\left(\frac{g^2}{8\pi^2}\right)^2\times\\
&&\left(\ln\left(\frac{g^2}{8\pi^2}\right)+\ln(\frac{2\pi^2N}{3})+\frac{3}{2}\right)
\Big\}.
\label{Gava}
\eeqa
With an appropriate running coupling, $g(T)$, this
defines the renormalized Polyakov loop up to ${\cal O}(g^4)$.
We make the specific choice of the two-loop formula,
\beq
   g^{-2}(T) = 2\beta_0\ln\left(\frac{\mu T}{\Lambda_{\bar{MS}}}\right)
    +\frac{\beta_1}{\beta_0} \ln\left(2\ln\left(\frac{\mu T}
          {\Lambda_{\bar{MS}}}\right)\right),
\label{2loop}
\eeq
with $\beta_0=11/(16\pi^2)$ and $\beta_1=102/(16\pi^2)^2$. 
and $T_c/\Lambda_{\overline{MS}}=1.14$.
\cite{Bali:1992ru,Karsch:2000ps,Gupta:2000hr}.
These predictions are shown in Figure \ref{fig:Lren_PGpaper2} for the
choices $\mu=\pi/2$, $\pi$ and $2\pi$. Due to the phase transition, the
Polyakov loop expectation value vanishes below $T_c$ and rises beyond
$5T_c$. It starts to decrease from about $10T_c$ and approaches the
asymptotic high temperature limit, $L_3^{r}=1$ (indicated by the arrow
in fig.~\ref{fig:Lren_PGpaper2}), from above, in qualitative agreement
with weak coupling theory. The lattice measurements seem to fall a
little slower than the HTL prediction, upto the highest temperature
examined. Approximate qualitative agreement with HTL perturbation theory,
without exact quantitative agreement upto very high temperature has been
seen in many other contexts in high temperature QCD, most notably for
screening masses \cite{Kaczmarek:1999mm, Datta:2002je} at high temperatures.\\
The comparison of the renormalization constants $Z_3(g^2)$ for different 
$N_\tau$, shown in Figure \ref{fig:renconst48} demonstrates
that $Z_3$ depends only on the bare coupling, and not on temperature.
The solid line in Fig.~\ref{fig:renconst48} shows the result of a fit with a
(two-loop) perturbation theory \cite{Heller:1984hx} inspired Ansatz,
\beq
Z_3(g^2) = \exp\left( g^2\frac{N^2-1}{N} Q^{(2)} + g^4 Q^{(4)}\right)
\label{z_pert}
\eeq
where $Q^{(2)}$ and $Q^{(4)}$ are expected to be independent of $N_\tau$
if eq.\ (\ref{pol_rel2}) is to be satisfied.
Although we are in coupling range which is not small enough for the weak coupling
expansion to be numerically accurate, 
the fit works surprisingly well. In fact the bare coupling becomes significantly
larger than unity before this ansatz begin to overestimate the actual values of
$Z_3$.
From the best fit analysis with a fit range $g^2\lsim 1.2$ we obtained the
values $Q^{(2)}=0.0591(21)$ and $Q^{(4)}=0.0605(54)$.
Interestingly, although our computations are done using a Symanzik improved
action, the value of $Q^{(2)}$ agrees reasonably well with
result from lattice perturbation theory, $Q^{(2)}=0.057(2)$ \cite{Heller:1984hx},
using the Wilson action.
The renormalization scheme dependence of these results
will be discussed in Appendix~\ref{sec:scale}.

\subsection{Polyakov loops in other representations}\label{sc.renorm}
\begin{figure}[thb]
\begin{center}
\scalebox{0.65}{\includegraphics{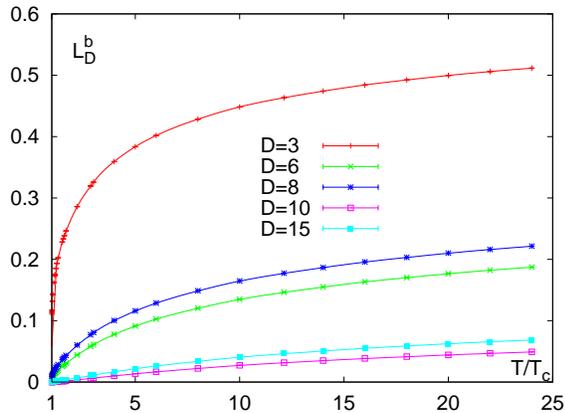}}
\end{center}
\caption{
The (bare) Polyakov loops for different representations $D$ measured on
$32^3\times 4$ lattices. The solid lines are the splines used in our analysis
to extract the renormalization constants in the direct renormalization procedure.
}
\label{fig:rpol}
\end{figure}
\begin{figure}[thb]
\begin{center}
\scalebox{0.65}{\includegraphics{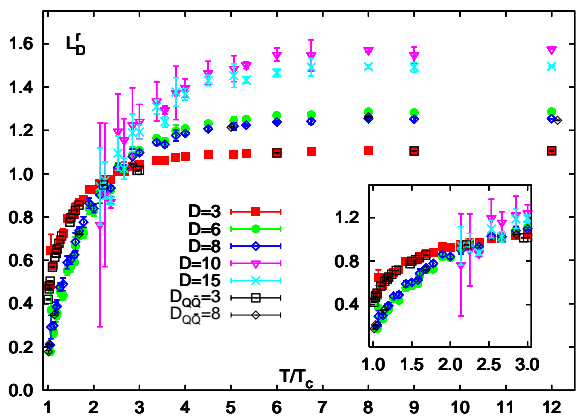}}
\end{center}
\caption{
The renormalized Polyakov loops for different representations $D$ obtained with
the direct renormalization procedure. Also shown are the results obtained from
the $Q\bar Q$-method for fundamental and adjoint loops, labeled $D_{Q\bar Q}$.
}
\label{fig:renpol2}
\end{figure}
The results for the measured bare Polyakov loops is shown in
fig.~\ref{fig:rpol} for the representations $D=3$ to $D=15$. 
The renormalization of Polyakov loops in each representation, $D$,
was obtained using the direct renormalization procedure. For each $D$,
the starting point was taken at $T_{ref}=12 T_c$. We fixed the scheme
through the choice
\beq
L^{r}_{D}(T_{ref}) = \left(L^{r}_{3}(T_{ref})\right)^{d_D}.
\label{starting}\eeq
Except for this assumption at a single temperature,
the renormalization was performed independently at each $D$.\\
Since the loops were measured at arbitrary temperature values,
spline interpolations (solid lines) for the data sets were used in the
renormalization iteration. The errors on renormalized Polyakov loops and
renormalization constants were obtained through a jackknife analysis.
The accumulation of errors during iteration, the exponential $d_D$
in (\ref{pol_rel2}), and the larger statistical errors for higher
representations lead to large errors in the renormalization procedure
for representations higher than $D=8$ as one approaches $T_c$.\\
The results for $L_D^{r}(T)$ are shown in Figure \ref{fig:renpol2}. For
comparison we have also included in the figure the results for the
fundamental and adjoint representation obtained with the $Q\overline Q$
procedure (see Appendix \ref{sec_plc8}).  These agree within errors, demonstrating again that the two
renormalization procedures give equivalent results.

\section{Casimir scaling}\label{sec:casimir}
\begin{figure}[thb]
\begin{center}
\scalebox{0.65}{\includegraphics{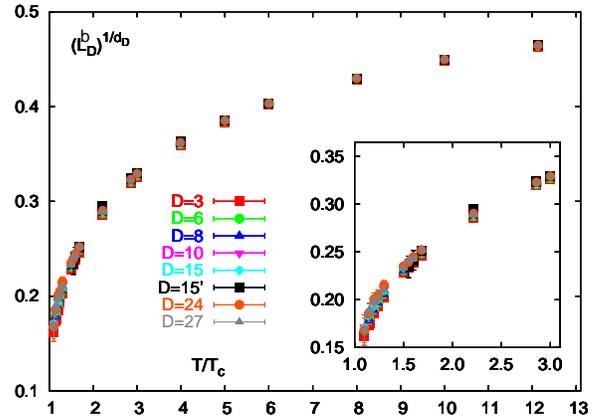}}
\end{center}
\caption{
The Casimir-scaled bare Polyakov loops for different representations $D$
measured on $32^3\times 4$ lattices.
}
\label{fig:pol_bare_T}
\end{figure}
\begin{figure}[thb]
\begin{center}
\scalebox{0.65}{\includegraphics{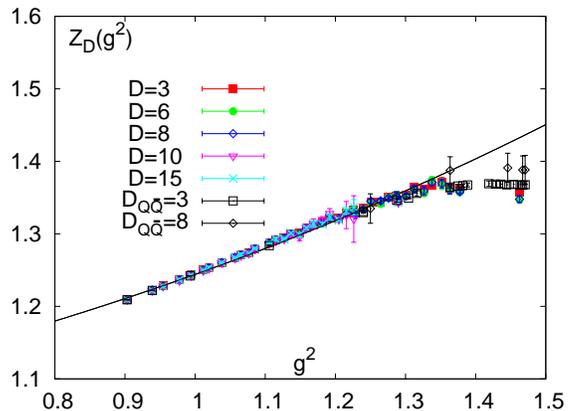}}
\end{center}
\caption{
The renormalization constants for different representations $D$ obtained with
the direct renormalization procedure. Also shown are the results obtained from
the $Q\bar Q$-method for fundamental and adjoint loops.
}
\label{fig:renconst}
\end{figure}
In \cite{Schroder:1998vy} it was shown that Casimir scaling is realized in
perturbation theory (at least) up to two-loop order, ${\cal O}(g^4)$. This
statement even holds for QCD with (massless) dynamical quarks as shown
in lattice perturbation theory in \cite{Bali:2002wf}.  Moreover, lattice
calculations at finite temperature employing an effective action for the
Polyakov loop in SU(3) have found Casimir scaling to be realized for the
Polyakov loop as well \cite{Damgaard:1987wh}.  Numerical calculations
on the lattice at $T=0$ in SU(3) pure gauge theory show that Casimir
scaling is realized also in the non-perturbative regime for distances
smaller than the string breaking distance \cite{Bali:2000un}.  The very
good agreement of the lattice data with the Casimir scaling hypothesis at
non-perturbative  distances in the vacuum has considerable ramifications
on models for non-perturbative QCD, especially for the confinement
mechanism \cite{Shevchenko:2000du}.\\
We have noted before that if Casimir scaling holds, then it holds for bare
as well as renormalized loops. Since bare loops have smaller statistical
errors, we test Casimir scaling through these. The most straightforward
test is to note that $\left(L^{b}_D\right)^{1/d_D}$ must be independent of
$D$ if Casimir scaling holds. In Figure \ref{fig:pol_bare_T} we show that
deviations from Casimir scaling are visible only very close to $T_c$. This
has implications for weak coupling expansions. Beyond two-loop order in a
perturbative series of the Polyakov loop, Casimir scaling violations
can appear \cite{Schroder:1998vy}.  These must be strongly suppressed
compared to contributions which scale with the quadratic Casimir operator.\\
An equivalent test is to note that the quantities $Z_D(g^2)$ which are
obtained using the assumption in eq.\ (\ref{starting}) are equal within
errors, as shown in Figure \ref{fig:renconst}. Note that this agreement
is an outcome of the renormalization procedure, and not built into it.
\begin{figure}[thb]
\begin{center}
\scalebox{0.65}{\includegraphics{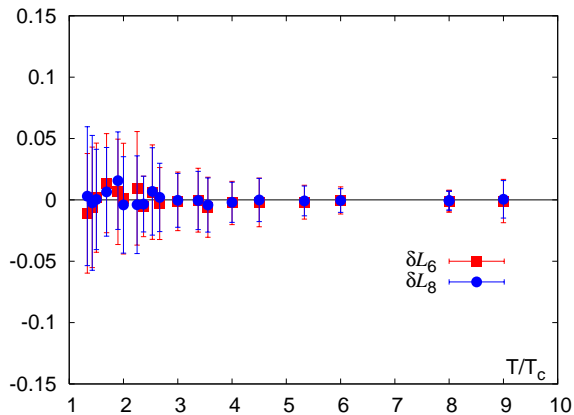}}
\end{center}
\caption{
Difference loops for the sextet, $\delta l_6$, and adjoint, $\delta l_8$,
Polyakov loops using Casimir scaling (\ref{eq_casimir}).
}
\label{fig:diff_loop2}
\end{figure}
\begin{figure}[thb]
\begin{center}
\scalebox{0.65}{\includegraphics{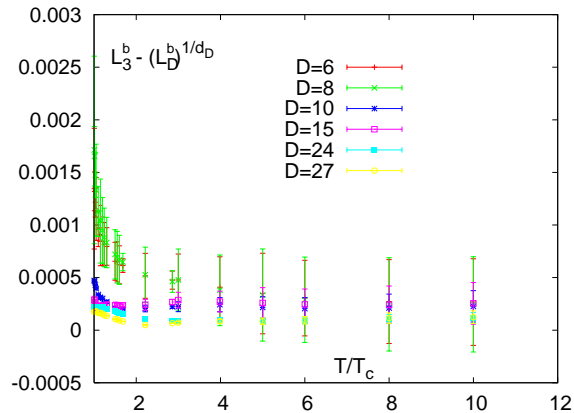}}
\end{center}
\caption{
Difference loops for all representation, $\delta l_D$,
using Casimir scaling (\ref{eq_casimir}).
}
\label{fig:rpol_diff}
\end{figure}
A finer test of Casimir scaling is obtained using the difference loops
\beq
   \delta L_D = L_3 - \left(L_D\right)^{1/d_D},
\label{eq_casimir}
\eeq
The results for the renormalized difference loops for $D=6$ and 8 are
shown in Figure \ref{fig:diff_loop2}. They are consistent with zero 
at all temperatures. For the higher representations a statistically
finer test is obtained with bare loops, since the errors on the
renormalized loops are large. The results are shown in
Figure \ref{fig:rpol_diff}. Even here, Casimir scaling is a good
approximation, which gets better the higher the temperature.
\subsection{Two-flavor QCD}
\begin{figure}[thb]
\begin{center}
\scalebox{0.65}{\includegraphics{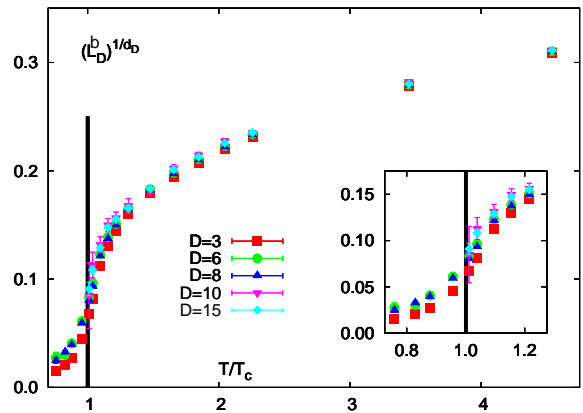}}
\end{center}
\caption{
Testing Casimir scaling for the bare Polyakov loop 
in 2-flavor QCD $\left\langle L_D\right\rangle$     
from a $16^3\times 4$ lattice for all $D=3,6,8,10,15$.
}
\label{fig:pol_t_fqcd}
\end{figure}
It is interesting to check whether Casimir scaling also holds for
QCD with quarks. Since dynamical
quarks break the center symmetry explicitly, and $N_f=2$ QCD has a
finite temperature cross over rather than a true phase transition,
the thermodynamic limit of the Polyakov loop below $T_c$ is non-vanishing.\\
In Figure \ref{fig:pol_t_fqcd} we show
$\left(L^{b}_D\right)^{1/d_D}$ for
$D=3,6,8$ at all temperatures, and $D=10,15$
above $T_c$. The latter representations are too noisy below $T_c$
to add any information. These scaled quantities are almost independent of $D$
down to $\sim 1.5T_c$. Below this
temperature we observe deviations to smaller values for the fundamental
representation, whereas the values for higher $D$ still coincide within errors.
Therefore we see a violation of Casimir scaling, between the fundamental
and other representations, when entering the
transition region which continues to the smallest temperatures analyzed.
These violations are relatively mild. Differences
between $(L^{b}_D)^{1/d_D}$ 
for $D=6$ and $D=8$ remain statistically insignificant even at
the smallest temperatures, as shown in the
inset of Figure \ref{fig:pol_t_fqcd}.

\section{The large-$N$ limit}\label{sec:large_nc}
\begin{figure}[thb]
\begin{center}
\scalebox{0.65}{\includegraphics{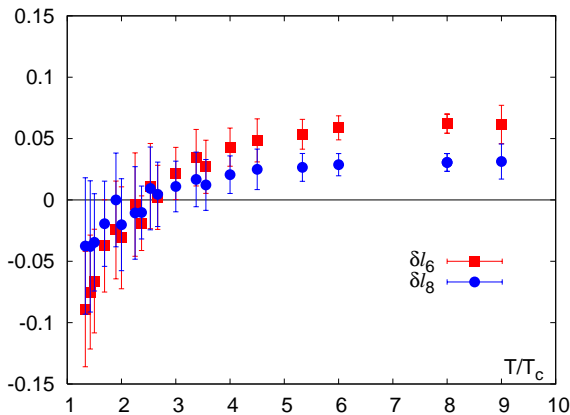}}
\end{center}
\caption{
Differences loops for the sextet, $\delta l_6$, and adjoint, $\delta l_8$.
}
\label{fig:diff_loop}
\end{figure}
The relation between Polyakov loops in different representations becomes rather
simple in the limit of a large numbers of colors, $N$ \cite{Dumitru:2003hp}.
In this limit the expectation value of a Polyakov loop in representation $D$
can be expressed in powers of the fundamental ($ L_{N}$) and anti-fundamental,
($L_{\overline{N}}=(L_{N})^*$) loop,
\beq
   L_D = L_{N}^{p_+} L_{\overline{N}}^{p_-} + {\cal O}\left(\frac1N\right)
\label{eq_largeN}
\eeq
where the integers $p_+$ and $p_-$ are determined from the Young tableaux
of the representation $D$. We investigate this large-$N$ factorization
using our data obtained for $N=3$.\\
Following \cite{Dumitru:2003hp} we analyze difference loops
\beqa
\delta l_6 &=& L^{r}_6 - L^{r}_3 \\
\delta l_8 &=& L^{r}_8 - \vert L^{r}_3 \vert^2.
\label{eq_pisarski}
\eeqa
Naively, the correction terms are expected to be of the order
of $(L^{r}_3)^2/3$, \ie, about 33\%.  Results, shown in Figure
\ref{fig:diff_loop} are clearly non-zero, except at around $2.5~T_c$
where all loops are one. Our results are comparable in magnitude to
those in \cite{Dumitru:2003hp} but show rather different temperature
dependence. Below $2 T_c$ the corrections are relatively large, and
the usefulness of the large-$N$ approximation seems doubtful. However,
above this temperature, the difference loops are of order 5--10\% of
the loop itself, and therefore significantly smaller than the naive
expectations. The large-$N$ approximation seems to fare better than
expected. This is similar to the conclusion reached for the equation of
state in \cite{Gavai:2004se}.

\section{Matrix Models}\label{matrixmodel}
One could seek effective field theories for Polyakov loops in the form
of matrix models, \ie, models in which each spatial site on the lattice,
$\vec x$, contains a matrix valued ``spin'', $l(\vec x)$,
\beq
   Z=\int\prod_{\vec x} dl(\vec x) \exp[-S_{MM}],
\label{partmm}\eeq
and the integration measure is the Haar measure. For $SU(N)$ gauge groups
the matrix takes values in SU(N). The action for these models can be
written in the form
\beqa
\nonumber
   S_{MM} &=& -\frac{N^2}d\sum_{DD'} \biggl[\beta_{D,D'}
      \delta_{t(D\times D'),0}\\\nonumber
      && \times\sum_{\vec x\hat n}
             \re l_D(\vec x) l_{D'}(\vec x+\hat n)\biggr.\\
       &&+\gamma_D\delta_{t(D),0}
     \sum_{\vec x} \re l_D(\vec x)\biggr],
\label{actionmm}\eeqa
where $\vec x$ runs over every site in the lattice, $\hat n$ over the
$2d$ nearest neighbors, $l_D$ is the Polyakov loop in the irreducible
representation $D$, i.e.
the trace of the matrix, and $t(D)$ is the triality of the irreducible
representation $D$. 
In this section we use the notation of Appendix \ref{su3}, i.e. traces are normalized to
the dimension of the corresponding representation, $D$. We also use
the notation $\ell_D=\langle l_D\rangle$.
The
constraint of vanishing triality arises from the $Z_N$ center-invariance
of the pure gauge theory.

For SU(3), the effective action with only the leading term
$\beta_{3,\overline 3}$ has been investigated extensively over the years.
However, when adding all irreducible representations upto a certain $D$, as $D$ varies, one
needs the couplings
\beqa
\nonumber
   \mathbf 3 &\quad& \beta_{3,\overline 3}\\
\nonumber
   \mathbf 6 &\quad& \beta_{3,6},\,\beta_{6,\overline 6}\\
\nonumber
   \mathbf 8 &\quad& \beta_{8,8},\,\gamma_8\\
\nonumber
   \mathbf{10} &\quad& \beta_{8,10},\,\beta_{10,10},\,\gamma_{10}\\
\nonumber
   \mathbf{15} &\quad& \beta_{3,\overline{15}},\,\beta_{6,15}\\
   \mathbf{15'} &\quad& \beta_{3,\overline{15'}},\,\beta_{6,15'},\,
         \beta_{15,\overline{15'}},
\label{couplings}\eeqa
and so on. \\
A matrix model would be used to obtain the loop expectation values
\beq
   \langle l_D^{r}(\{\beta\})\rangle = \frac1Z
     \int\prod_{\vec x} dl(\vec x) l_D \exp[-S_{MM}(l,\{\beta\})],
\label{mmexpect}\eeq
where $\{\beta\}$ denotes the whole set of couplings in the action.
Equating these expressions to a sufficient number of observations
on $\ell_D^{r}(T)$, one would obtain the temperature
dependence of the couplings. Other predictions of matrix models,
which we do not explore here, are expectation values of moments (for
example, the Polyakov loop susceptibilities) and correlation functions
of loops.\\
Note an intrinsic complication in the matching procedure. Since
$\ell_D^{r}(T)$ is scheme dependent, 
the couplings that one extracts by any matching procedure must also
be scheme dependent.  Note also that, in order to make contact with
a matrix model, one has to choose a renormalization scheme in which
$\ell_D^{r}(T)<D$.  To the best of our knowledge, these
points have not been noted in the literature.
\begin{figure}[tb]
\begin{center}
\scalebox{0.65}{\includegraphics{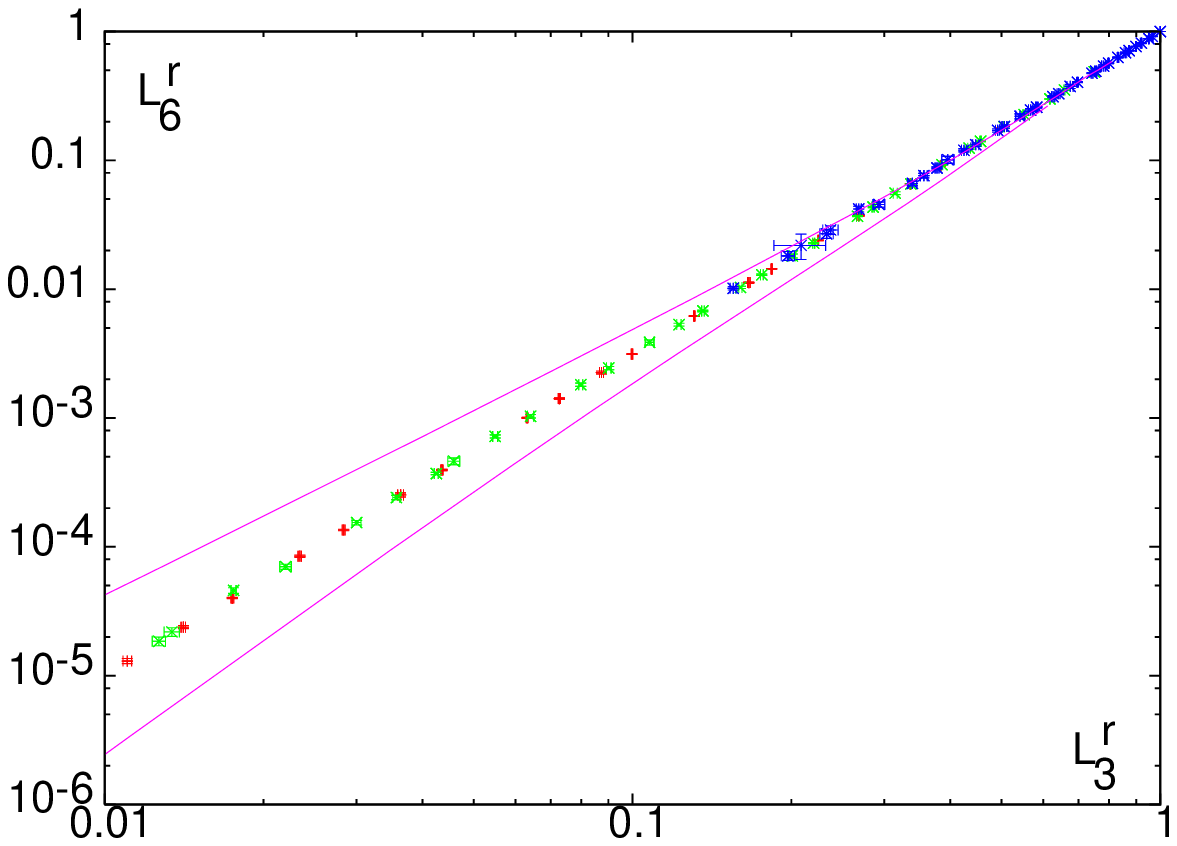}}
\scalebox{0.65}{\includegraphics{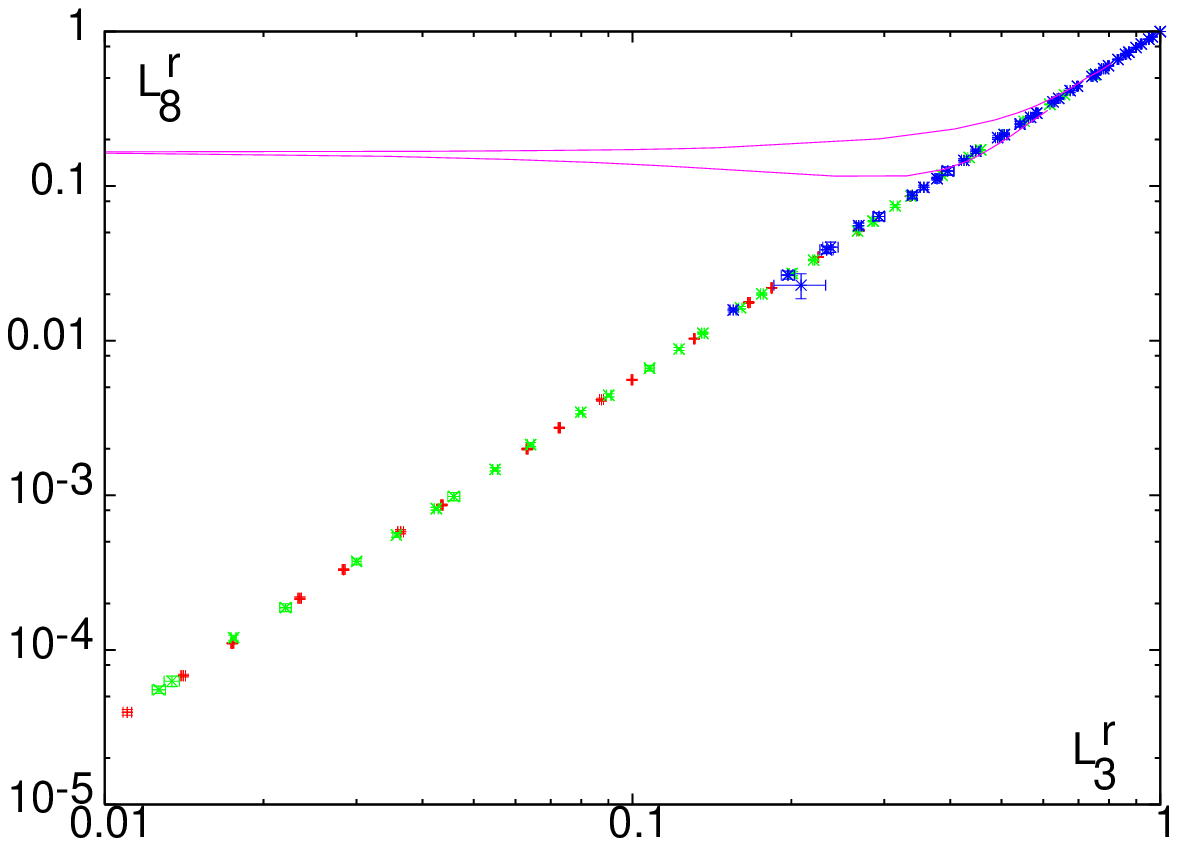}}
\end{center}
\caption{Different renormalized Polyakov loops shown as a function of
 the fundamental loop. The fact that the data in several different
 renormalization schemes ($\ell_3^{r}(12T_c)=0.5$ in red,
 0.75 in green and 1 in blue) collapse on to a single universal curve in
 each case implies that there can be only a single coupling matrix model
 which describes this data. The line is the result of a fit using four
 terms in the action, as described in the text, tuned to bracket the
 observed curve $\ell_6(\ell_3)$.}
\label{fig:mmuniv}\end{figure}
Furthermore, as the number of irreducible representations increases, the number of couplings
in the effective theory which need to be matched to data increases
rapidly. A determination of the effective action involves extraction
of the couplings through such matching at each temperature.  This is an
ill-conditioned problem unless the series can be cut off, and the number
of couplings required is less than the number of pieces of data.\\
It is interesting to ask how one can bound the number of couplings needed
in the matrix model. If there are $C_N$ couplings to be determined,
then $C_N$ of the expectation values can be traded for the couplings,
and all other expectation values can be written in terms of these. For
example, for a one-coupling matrix model, $C_N=1$, one could write
$\ell_D^{r}(\ell_3^{r})$, for all $D>3$. This
relation is RG invariant: in two different renormalization schemes, if
the values of $l_3^{r}$ at two different temperatures are the same,
then the values of $l_D^{r}$ will be equal, for each $D$.  For the pure
gauge theory the data in Figure \ref{fig:mmuniv} shows that the SU(3) pure
gauge theory requires a matrix model with only a single coupling.  A single
coupling matrix model means that ratios such as $\beta_{3,6}/\beta_{8,8}$
are fixed, and only one coupling is dependent on the temperature.\\
The temperature independent ratios of couplings define the shape of the
universal curves, $\ell_D^{r}(\ell_3^{r})$,
and the single tunable coupling says how the curve is traversed, in a
given renormalization scheme, as $T$ changes. Therefore, one can solve
the problem in two steps: first use the universal curves to fix the ratios
of the couplings, and finally solve the easier problem of finding the
single left over coupling. Since exact solutions for the loop expectation
values are not known for matrix models with $N_c=3$, one has to either
solve the problem through a Monte Carlo simulation or in mean field theory.
Here we investigate the latter option.\\
Taking into account the irreducible representations $\mathbf 3$ and $\mathbf 6$ in the effective
action, one has
\beqa
\nonumber
   S &=& -3\sum_{\vec x,\hat n} \biggl[
     \beta_{3,\overline3}\re l_3(\vec x) l^*_3(\vec x+\hat n)\nonumber\\
   &&\quad\biggl. + \beta_{3,6}\re l_3(\vec x) l_6(\vec x+\hat n)\biggr.\nonumber\\
   &&\quad\biggl.
     +\beta_{6,\overline6}\re l_6(\vec x) l^*_6(\vec x+\hat n) \biggr],
\label{seff}\eeqa
where $l_6=(l_3)^2-l_3^*$. Using this SU(3) relation, and making a mean-field
approximation, we find that
\beqa
\nonumber
   S &=& -6dV \biggl[
     \beta_{3,\overline3}\ell_3\re l_3 +
     \frac{\beta_{3,6}}2\{\ell_6\re l_3 + \ell_3\re (l_3^2-l_3^*)\}\biggr.\\
     &&\quad\biggl.
     +\beta_{6,\overline6}\ell_6\re (l_3^2-l_3^*)\biggr],
\label{smft}\eeqa
It is also possible to extend such a mean field treatment
to models which include the octet representations. Using the invariance of
the Haar measure, we can diagonalize the matrix, so that $l_3=\exp(i\phi)
+ \exp(i\psi) + \exp[i(\phi+\psi)]$, and perform the integration over
the remaining variables to give
\beqa
\nonumber
   dl_3 &=& \frac1{3\pi^2}\{1-\cos(\phi-\psi)\} \{1-\cos(2\phi+\psi)\}\\
     &&\qquad\times\{1-\cos(\phi+2\psi)\} d\phi d\psi.
\label{haar}\eeqa
Putting all this together, we find
\beqa
\nonumber
   Z(\beta_{3,\overline3},\beta_{3,6},\beta_{6,\overline6},\ell_3,\ell_6) &=&
      \left[\int dl_3\exp(-S/V)\right]^V\\
    &=& \exp[-VF].
\label{partmft}\eeqa
In the mean-field theory the expectation values are computed simply as
\beq
   \ell_D = \frac1Z \left[\int dl_3\exp(-S/V)\right] \re l_D,
\label{expectvals}\eeq
where the $l_D$ can be expressed in terms of the angles $\psi$ and $\phi$
using the formulae in eqs.\ (\ref{pol1}-\ref{pol7}) and the relation $\re l_3 = \cos\psi
+\cos\phi+\cos(\psi-\phi)$.\\
Some of the results are shown in Figure \ref{fig:mmuniv}. The observed
Casimir scaling of loops implies a power-law dependence of loops on
each other. The matrix
model which includes only the coupling $\beta_{3,\overline 3}$ is
in fair agreement with the universal curve $\ell_6(\ell_3)$ when
$\ell_3>1/2$. However, it disagrees with the curve for
$\ell_8(\ell_3)$ already when $\ell_3=0.9$.  By including terms in
$\beta_{3,6}$ and $\beta_{6,\overline6}$ (in a fixed $T$-independent ratio
to $\beta_{3,\overline3})$ the curve for $\ell_6$ can be improved; but
this leads to no perceptible change in the curve for $\ell_8$. However,
by introducing the coupling $\beta_{8,\overline8}$, and tuning the
$T$-independent ratio $\beta_{8,\overline8}/\beta_{3,\overline3}$, one can
contrive to improve the description of the two universal curves. However
the universal curves for $\ell_{10}$ \etc, need further tuning.\\
The conclusion seems robust: the SU(3) pure gauge theory data can
be described within a single coupling matrix model. However, it seems
hard to construct a matrix model with a small number of terms which
reproduces the power-law dependence of $\ell_D$ on $\ell_3$. The second
result has been obtained within a
mean field theory, and needs verification in a more complete approach,
such as the full simulation of such matrix models.

\section{Adjoint sources and gluelumps}
\label{sec_strinbreaking}

\begin{figure}[t]
\begin{center}
\scalebox{0.68}{\includegraphics{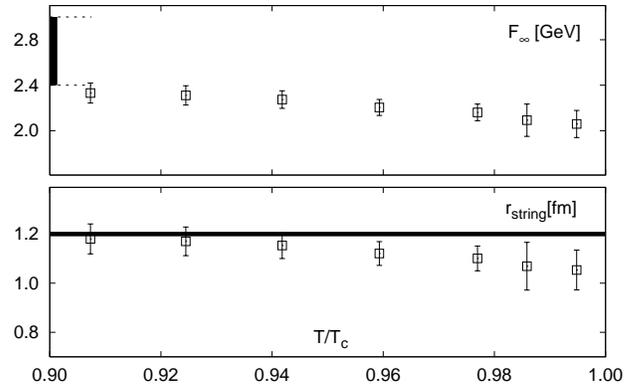}}
\end{center}
\caption{
Asymptotic values of the adjoint heavy quark free energies (upper
panel). Estimates of the string-breaking radius using the two definitions
explained in the text (lower panel).
}
\label{fig:mglump_rstring}
\end{figure}

Polyakov loops in representations with non-zero triality vanish in the
confined phase of the pure gauge theory, since the Z(3) symmetry of
the action is realized on the states with large contribution to the
path integral. The behavior of loops with vanishing triality can be
different, because they are blind to the Z(3) symmetry involved in the
QCD phase transition.  This study is confined to the octet loop since
all other triality zero loops that we constructed turned out to have
very large errors below $T_c$.\\
\begin{figure}[thb]
\begin{center}
\scalebox{0.65}{\includegraphics{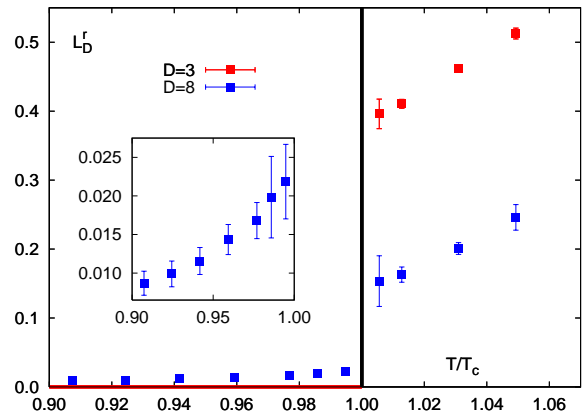}}
\end{center}
\caption{
The renormalized adjoint Polyakov loops below and above the critical
temperature compared to the fundamental loops above $T_c$. The solid line below
$T_c$ indicates the vanishing fundamental loops in the confined phase. 
}
\label{fig:pol_adj2}
\end{figure}
A dynamical picture has been
advanced for the behavior of adjoint Polyakov loops below $T_c$.  An adjoint source can couple to a gluon in the medium to
form a colorless composite called a gluelump. Correlations of triality
zero loops can clearly be mediated by gluon exchange at any temperature,
leading to screening. Gluelumps provide a summary of the main features
of such screening \cite{Karsch:1998qj} through two parameters: the free
energy of separated gluelumps determines the asymptotic value of an octet
Polyakov loop, the string breaking distance is the distance at which the
long-distance screening behavior sets in.\\
We found that the bare adjoint Polyakov loop $L^{b}_8$ below $T_c$
assumes its thermodynamic limit, \ie, becomes independent of the volume
for $N_\sigma^3\times 4$ lattices with $N_\sigma \ge 24$.  $L^{b}_8$ could
be renormalized using either the $Q\overline Q$ method for the octet loop
explained in Appendix \ref{sec_plc8}.  However, we calculated $L^{b}_8$
at more couplings than the adjoint correlator. Since we found that $Z_8$
agrees with $Z_3$ (see Appendix \ref{sec_plc8}), we used the $Z_3(g^2)$
given in Table \ref{table_fz} to obtain $L^{r}_8(T)$.\\
Table \ref{table_f8} lists the values found for the renormalized
adjoint Polyakov loop $L^{r}_8$ and Figure \ref{fig:pol_adj2} compares
these results to those for the renormalized fundamental Polyakov loop,
$L^{r}_3$.  Although $L^{r}_8$ becomes rather small below $T_c$, it is
clearly non-zero for all temperatures analyzed by us.  We observe that
$L^{r}_8$ rises from $0.0087(16)$ at $T/T_c=0.907$ to $0.0219(48)$ just
below $T_c$ at $T/T_c=0.995$. Crossing the critical temperature into
the deconfined phase,  $L_8^{r}$ jumps almost an order of magnitude to
$0.154(37)$ at $T/T_c=1.005$. It is a little surprising to find the octet
loop, which is blind to the Z(3) symmetry, change discontinuously at the
symmetry-breaking transition. However, other triality-zero operators also
change discontinuously at the phase transition, most notably the energy
density.
\begin{figure}[tbh]
\begin{center}
\scalebox{0.65}{\includegraphics{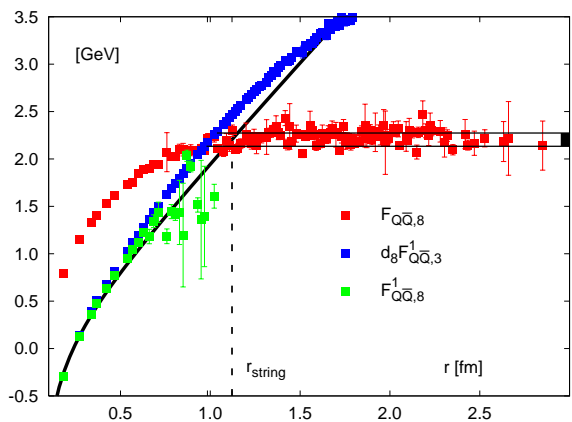}}
\end{center}
\caption{
Heavy quark-antiquark free energies for adjoint sources in the color singlet
and color averaged channel compared to the Casimir scaled color singlet free
energy of fundamental sources at a temperature of $0.959\,T_c$. 
The dashed line indicates the definition of
the string-breaking radius as explained in the text and the horizontal 
solid lines show the
asymptotic value for the adjoint free energies. 
The thick black line indicates the adjoint $T=0$ potential $V_8(r)=d_8V_3(r)$.  
}
\label{fig:pol_free_adj}
\end{figure}
We now address the issue of string breaking and determine the binding energy
of the gluelump. The specific situation at $T/T_c=0.959$, shown in
Figure \ref{fig:pol_free_adj} serves as an example. We have shown
$F^1_{Q\bar Q,8}$, and, since it becomes too noisy at large distances,
also the ``color average'' free energy $F_{Q\bar Q,8}$, which has the
same value at long distances. These free energies clearly show that
adjoint sources are screened at large distances, in clear contrast to the
linearly rising free energy, $F_{Q\bar Q,3}$ of sources in the
fundamental representation. We calculate the free energy at infinite
separation between the sources using the cluster property,
\beq
F_{8,\infty}(T) = -2T\ln L_8^{r}(T) = 2 m_{\text{glump}}(T).
\label{eq:f_inf_L8} 
\eeq
At zero temperature the energy stored in the field suffices to
put on-shell two gluons from the medium and form two disjoint
gluelumps \cite{Bali:2003jq}. At finite temperature this
free energy can be identified with twice the gluelump screening mass.\\
Results for $F_{8,\infty}$ are collected in \ref{table_f8}.
$F_{8,\infty}$, is shown
in Figure \ref{fig:mglump_rstring} (upper panel).
It changes
little with $T$, starting from $2.331(88)$ GeV at $T/T_c=0.907$ and
subsequently falling to $2.06(12)$ GeV just below $T_c$.
At the lowest temperatures discussed here, $F_\infty$ indeed approaches
the $T=0$ value of $V_{8,\infty}=2.4 - 3.0$ GeV, which is twice the
mass of the gluelump obtained in \cite{Simonov:2000ky}.\\
We define the string breaking distance $r_{\text{string}}(T)$ by comparing
the Casimir scaled free energy with fundamental sources, $d_8 F^1_{Q\overline Q,3}$,
with the screened value of the free energy with adjoint sources,
\beq
d_8 F^1_{Q\overline Q,3}(r_{\text{string}}(T))=F_{8,\infty}(T).
\label{eq:r_string_V8}
\eeq
\begin{table}[thbp]
  \begin{center}
    \begin{tabular}{|c|l|c|c|}
      \hline
     $T/T_c$ & $F_\infty$ [GeV] & $r_{\text{string}}(V_8)$ [fm] & $L_8^{r}$\\
     \hline
     $0.907$ &  $2.331(88)$ &  $1.180(61)$ &   $0.0087(16)$ \\
     $0.924$ &  $2.310(84)$ &  $1.170(58)$ &  $0.0099(17)$\\
     $0.942$ &  $2.274(76)$ &  $1.153(53)$ &   $0.0116(17)$\\
     $0.959$ &  $2.204(70)$ &  $1.121(48)$ &   $0.0143(19)$\\
     $0.977$ &  $2.161(73)$ &  $1.101(50)$ &   $0.0168(23)$\\
     $0.986$ &  $2.09(14)$ &  $1.069(97)$ &   $0.0198(53)$\\
     $0.995$ &  $2.06(12)$ &  $1.053(81)$ &   $0.0219(48)$\\
     \hline
      $0.000$  & $2.4-3.0$ & {$\sim 1.2$}  & -- \\
     \hline
    \end{tabular}
    \caption{\label{table_f8}Temperature dependence of $F_\infty$, string breaking distance
      $r_{\text{string}}$ for the adjoint singlet free energy
      with respect to $V_8$ (see text) and the renormalized adjoint
      Polyakov loop $L_8^{r}$.
      The last line gives the values at $T=0$ for twice the mass of the
      gluelump \cite{Simonov:2000ky} and for
      the string breaking distance \cite{Michael:1998sm}.
    }
  \end{center}
\end{table}

The results are collected in Table \ref{table_f8}.
In Figure \ref{fig:mglump_rstring} (lower panel) we show the resulting
values of $r_{\text{string}}$ as a function of the temperature. There
is rather mild change in $r_{\text{string}}$ 
with $T$. It varies from $1.180(61)$ fm at $T/T_c=0.907$ 
to $1.053(81)$ fm just below $T_c$.
At the smallest temperature $r_{\text{string}}$ almost coincides with
the $T=0$ value of $1.2$ fm \cite{Michael:1998sm}.

\section{Conclusions}

We examined the renormalized Polyakov loop in many different irreducible
representations of the gauge group SU(3) in the thermodynamic limit of
pure gauge QCD. It has been known for a long time that the ultraviolet
divergences of the Polyakov loop can be absorbed into a multiplicative
renormalization ``constant'' $Z(g^2)$, where $g^2$ is the bare coupling.
Such a renormalization factor does not depend on long distance physics,
such as the temperature, $T$ (see eq.\ \ref{pol_rel2}). We implemented
such a renormalization procedure by explicitly constructing an iteration
using only explicitly gauge invariant quantities starting from a
reference temperature $T_{ref}$ incorporating this idea (see Section
\ref{sec_direct} for details). This so-called direct renormalization
procedure was then used to extract the renormalized Polyakov loops
in representations upto the {\bf 27} of SU(3) for a wide range of
temperatures (see Section \ref{sec:sim}).

The technical part of our paper also consists of extending the $Q\overline
Q$ renormalization procedure of \cite{Kaczmarek:2002mc} to Polyakov
loops in arbitrary representations of the gauge group (see Appendix
\ref{sec_plc8}). This is done by matching (gauge variant) correlation
functions of sources in arbitrary representations to zero temperature
values at the ultraviolet cutoff. Although one does not demand explicitly
that the renormalization constant depends only on the bare coupling,
the matching to zero temperature in the ultraviolet makes sure that this
occurs. We checked that both renormalization procedures have the same, one
real parameter, freedom of choice of scheme (see Appendix \ref{sec:scale}).
Having two drastically different renormalization schemes which are
functionally equivalent allows us not only to use the most convenient
scheme in any situation, but also to cross check the results by using
both schemes whenever possible. This puts the results of the lattice
measurements on very strong footing. Furthermore, the equivalence of
the two procedures shows that the short distance as well as the large
distance parts of the heavy quark free energies obtained in Coulomb
gauge become gauge independent as proposed in \cite{Philipsen:2002az}.

An interesting simplification occurs when Polyakov loop
expectation values satisfy a relation called Casimir scaling (eq.\
\ref{casimirscaling}). Then the renormalization factors in all the
different representations essentially boil down to a single factor.
Furthermore, large-$N$ factorization evolves from the large-$N$ limit
of the quadratic Casimirs, and hence Casimir scaling could provide
an alternative route to large-$N$ scaling.
We have presented tests of Casimir scaling in Section \ref{sec:casimir}
and of direct large-$N$ factorization in Section \ref{sec:large_nc}. Both
turn out to be reasonably reliable away from $T_c$. However, Casimir
scaling is significantly more reliable and may provide a good route to
scale large-$N$ predictions down to $N=3$.

One subject of abiding interest is whether the high temperature phase of
QCD can be described by a matrix model. We test this question in Section
\ref{matrixmodel}. Casimir scaling implies that there are universal
(renormalization scheme independent) relations between the renormalized
Polyakov loop expectation values such that all the loops we studied depend
only on the value of the fundamental loop. This implies that a matrix
model description could work well away from $T_c$. A single parameter
variation of all couplings in the model would then reproduce the data
on Polyakov loops, the temperature dependence of the couplings being,
of course, renormalization scheme dependent.  However, it seems that a
simple model with a small number of parameters is not able to reproduce
the power-laws in the lattice data, at least within the mean-field
analysis of the matrix model performed here.

Due to the Z(3)-symmetry of the pure gauge theory, all Polyakov loops with
non-zero triality vanish in the confined phase of the pure gauge theory.
For the adjoint representation we have observed small, but non-zero,
values below $T_c$ for the first time in the thermodynamic limit (see
Section \ref{sec_strinbreaking}).  Since static adjoint sources can form
bound states, called gluelumps, with dynamical gluons, correlations
of adjoint loops show screening (string breaking) even in the confined
phase. As a result, heavy quark free energies have a finite asymptotic
value while for zero-zero triality they rise linearly with distance. Some
aspects of the free energy can be captured into the phenomenology of
gluelumps through a mass and radius parameter.  We present results for
these quantities.

Our primary technical result is the systematic development of two parallel
renormalization procedures for Polyakov loops in arbitrary representations
of the gauge group. This allows us to check that Casimir scaling of the
renormalized loops is satisfied to good accuracy away from $T_c$. This is
our main physical result, since it leads on to the discussion of large-$N$
factorization and the matrix model description of lattice data.

\section*{Acknowledgment}

We wish to thank J.~Engels, F.~Karsch, R.~D.~Pisarski, Y.~Schröder and F.~Zantow for
fruitful discussions.
This work has partly been supported by
contract DE-AC02-98CH10886 with the U.~S.~Department of Energy.
At an early stage of this work
K.~H.~ has been supported by the DFG under grant GRK 881/1.
S.~G. would like to acknowledge the hospitality of the University of
Bielefeld.

\bibliographystyle{h-physrev3} \bibliography{polrep}

\newpage

\appendix
\section{Polyakov loops in irreducible representations of SU(3)}
\label{su3}
In order to obtain Polyakov loops in higher irreducible representations of
SU(3) than the
fundamental, we may use the theorem that the character in a direct
product representation is the product of the corresponding characters,
$\chi_{P\times Q}(g)=\chi_P(g)\chi_Q(g)$. Then the direct product can
be reduced using the Clebsh-Gordan series to yield the Polyakov loop
in various representations.

We use the following identities:
\beqa
    \mathbf 3\times\mathbf 3 &=& \mathbf 6 +\overline{\mathbf 3}\nonumber\\
       (1,0)\times(1,0) &=& (2,0) + (0,1)\\
    \mathbf 3\times\overline{\mathbf 3} &=& \mathbf 8 +\mathbf 1\nonumber\\
       (1,0)\times(0,1) &=& (1,1) + (0,0)\\
    \mathbf 6\times\mathbf 3 &=& \mathbf{10} +\mathbf 8\nonumber\\
       (2,0)\times(1,0) &=& (3,0) + (1,1)\\
    \mathbf 6\times\overline{\mathbf 3} &=& \mathbf{15} +\mathbf 3\nonumber\\
       (2,0)\times(0,1) &=& (2,1) + (1,0)\\
    \mathbf 8\times\mathbf 3 &=& \mathbf{15} + \overline{\mathbf 6} + \mathbf 3\nonumber\\
       (1,1)\times(1,0) &=& (2,1) + (0,2) + (1,0)\\
    \mathbf{10}\times\mathbf 3 &=& \mathbf{15}' +\mathbf {15}\nonumber\\
       (3,0)\times(1,0) &=& (4,0) + (2,1)\\
    \mathbf{10}\times\overline{\mathbf 3} &=& \mathbf{24} +\mathbf 6\nonumber\\
       (3,0)\times(0,1) &=& (3,1) + (2,0)\\
    \mathbf 6\times\mathbf 6 &=& \mathbf{15}' +\mathbf{15}+\overline{\mathbf 6}\nonumber\\
       (2,0)\times(2,0) &=& (4,0) + (2,1) + (0,2)\\
    \mathbf 6\times\overline{\mathbf 6} &=& \mathbf{27} + \mathbf 8 + \mathbf 1\nonumber\\
       (2,0)\times(0,2) &=& (2,2) + (1,1) + (0,0)
\label{cgs}\eeqa
where we have specified the irreducible representations both in terms of its dimension
and through the canonical label $(p,q)$ where $p$ and $q$ are
integers. Recall that the maximum weight in irreducible representation $(p,q)$ is $\mathbf
m=\{(p+q)/2\sqrt3,(p-q)/6\}$, and the dimension of this irreducible representation is
$D=(p+1)(q+1)(p+q+2)/2$. The notation $\mathbf{15}'$ stands for the
irreducible representation $(4,0)$, and $\mathbf{15}$ denotes the irreducible representation $(2,1)$. Note
that interchanging $p$ and $q$ gives the complex conjugate irreducible representation. The triality of
an irreducible representation can be defined to be $t=(p-q)|_3$. In each expression above,
the trialities of all the irreducible representations on the right must be equal, and must equal
the sum of the trialities (mod 3) of the irreducible representations on the left. This can be
used as a check.\\
More concretely, take the product over links and the trace defining the
Polyakov loop in the $\mathbf 3$ of SU(3),
\beq
   l_{\mathbf 3}(x) = \tr \prod_{n=1}^{N_\tau} U_t(x+n\hat t),
\label{polythree}\eeq
where $l_{\mathbf 3}(x)$ is a complex number. Here the trace is normalized
such that the unit matrix traces to 3. The Polyakov loop in an
irreducible representation is the character in that irreducible representation. Hence, given the loop in one irreducible representation, that
in the complex conjugate irreducible representation is obtained by complex conjugation. One specific
example is
\beq
   l_{\overline{\mathbf 3}}(x) = l_{\mathbf 3}^*(x),
\label{polythreeb}\eeq
where $l^*$ is the complex conjugate of $l$. The Polyakov loop in the trivial
irreducible representation $\mathbf 1$ is unity (which gives vanishing potential in this irreducible representation).
Next we construct the series of other Polyakov loops,
\beqa
   l_{\mathbf 6}(x) &=& l_{\mathbf 3}(x)^2 - l_{\mathbf 3}^*(x),\label{pol1}\\
   l_{\mathbf 8}(x) &=& \left|l_{\mathbf 3}(x)\right|^2 -1,\label{pol2}\\
   l_{\mathbf {10}}(x) &=& l_{\mathbf 3}(x)l_{\mathbf 6}(x) - l_{\mathbf 8}(x),\label{pol3}\\
   l_{\mathbf {15}}(x) &=& l_{\mathbf 3}^*(x) l_{\mathbf 6}(x) - l_{\mathbf 3}(x),\label{pol4}\\
   l_{\mathbf {15}'}(x) &=& l_{\mathbf 3}(x) l_{\mathbf{10}}(x) - l_{\mathbf{15}}(x),\label{pol5}\\
   l_{\mathbf {24}}(x) &=& l_{\mathbf 3}^*(x) l_{\mathbf{10}}(x) - l_{\mathbf 6}(x),\label{pol6}\\
   l_{\mathbf {27}}(x) &=& \left|l_{\mathbf 6}(x)\right|^2 - l_{\mathbf 8}(x) -1.
\label{pol7}\eeqa
Further irreducible representations can be obtained if needed. Two of the reductions for
the direct product have not been used. 
The Polyakov loop values in normalization used elsewhere in this paper
is obtained by writing $L_D(x)=l_D(x)/D$.

\begin{table}
  \begin{center}
    \begin{tabular}{lccccc}
      &&&&&\\
      $D$ & $(p,q)$ & $t$ & $C_2(D)$&$d_D$&\\
      \hline
      $3$   & $(1,0)$ & 1 &$4/3$& $1$&  \\
      $\overline{3}$ & $(0,1)$ & 2 & $4/3$& $1$ &  \\
      $6$   & $(2,0)$ & 2 & $10/3$& $5/2$&\\
      $8$   & $(1,1)$ & 0 &$3$ & $9/4$& $\mathrm{Im}(L_8)=0$\\
      $10$  & $(3,0)$ & 0&$6$  & $9/2$&\\
      $15$  & $(2,1)$ & 1 &$16/3$ & $4$&\\
      $15^{\prime}$ & $(4,0)$ & 1 &$28/3$ & $7$&\\
      $24$ & $(3,1)$ & 2 & $25/3$ & $25/4$&\\
      $27$  & $(2,2)$ & 0 &$8$ & $6$& $\mathrm{Im}(L_{27})=0$\\
      \hline
      &&&&&\\
    \end{tabular}
    \caption{\label{tab:casimir}Quadratic Casimir $C_2(D)$ for the
      representation $D$ of SU(3), $t=p-q \mbox{ mod } 3$ is the
      triality. $d_D$ is the ratio $C_2(D)/C_2(3)$.}
  \end{center}
\end{table}
%




\begin{figure}[thb]
\begin{center}
\scalebox{0.65}{\includegraphics{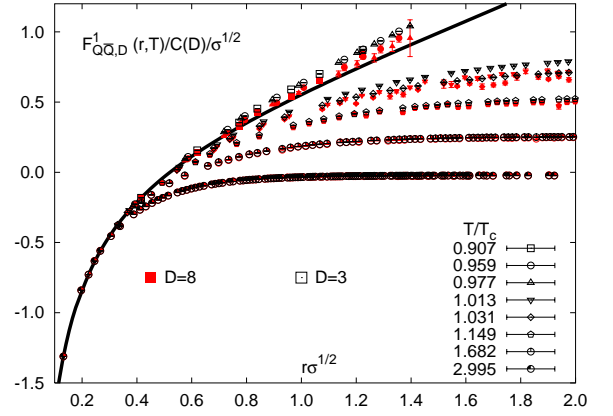}}
\end{center}
\caption{
Comparison of the color singlet quark-antiquark free energies for fundamental and adjoint
sources scaled by the corresponding Casimir factor. The solid line represents
the zero temperature potential, $V_8(r) = V_3(r)/C_2(8)$.
}
\label{fig:vgl_div}
\end{figure}

\begin{figure}[hbt]
\begin{center}
\scalebox{0.65}{\includegraphics{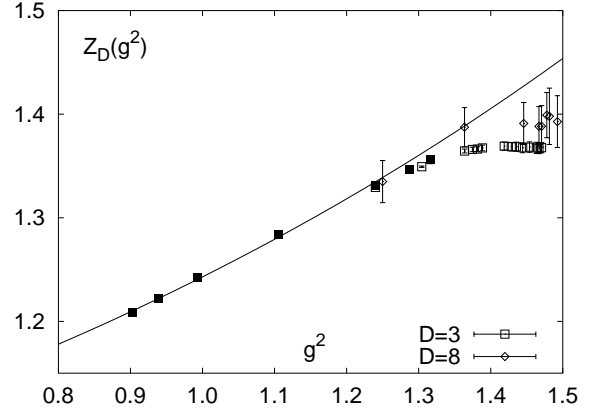}}
\end{center}
\caption{
The renormalization constants, $Z_D(g^2)$, for fundamental ($D=3$) and
adjoint ($D=8$) sources plotted vs. the bare coupling $g^2=2 N/\beta$. The solid
line is the same as in fig.~\ref{fig:renconst48}.
}
\label{fig:renconst38}
\end{figure}

\section{$Q\overline Q$ renormalization}\label{sec_plc8}

The $Q\overline Q$ renormalization procedure 
\cite{Kaczmarek:2002mc} can be extended to static
sources in arbitrary representations of the color group. For simplicity we will
only discuss adjoint sources here in detail, but the generalization to other
representations is straight forward.

Given an SU(3) matrix in the fundamental representation, $U^{3}$, the
corresponding adjoint matrix, $U^{8}$, is
\beq
  \label{adjoint_link}
  U^{8}_{ij} = \frac{1}{2}\tr \left(\lambda_i U^3 \lambda_j
  {U^3}^{\dagger} \right),\quad i,j=1,\dots,8,
\eeq
where $\lambda_i$ are the Gell-Mann matrices. From the hermiticity of
$\lambda_i$ and
cyclicity of the trace all matrix elements $U^{8}_{ij}$ are real. This
formula can be used to convert all link elements from the fundamental
to the adjoint.

The thermal Wilson line in the adjoint representation is
\beq
  \label{wilsonloop}
  P_8(x)=\prod_{x_4=0}^{N_\tau-1} U^8_4(\vec x,x_4).
\eeq
Another way to define it is to take the fundamental Wilson line of
eq.\ (\ref{wilsonline}) and convert it to the adjoint using the prescription
of eq.\ (\ref{adjoint_link}).
The adjoint Polyakov loop is the trace
\beq
L_8(\vec x) = \tr P_8(\vec x).
\eeq
As before, we have normalized the trace such that the trace of the
unit matrix is 1.

Define the correlator of two adjoint thermal Wilson lines,
\beq
\tilde C^1_{Q\bar Q,8}(r,T) = \left\langle \tr \left( P_8(\vec x_1)
    P^\dagger_8(\vec x_2) \right) \right\rangle,
\label{c1qq}
\eeq
and $r=\vert \vec x_1 - \vec x_2 \vert$. This correlator is clearly gauge
dependent, and hence we define it through Coulomb gauge fixing (see
\cite{Philipsen:2002az, Jahn:2004qr} for more on this point). The free
energy with two static adjoint sources a distance $r$ apart is
\beq
\label{eq:free_energy}
\tilde F^1_{Q\bar Q,8}(r,T) = -T \ln \tilde C^1_{Q\bar Q,8}(r,T).
\eeq
Here $\tilde C$ and $\tilde F$ denote bare correlators and
free energies; the same notation without a tilde will denote
renormalized quantities. 

Since the Polyakov loop is renormalized multiplicatively, the free energies are
additively renormalizable. We match the finite temperature free
energy to the zero temperature potential at the smallest attainable distance,
$a$, on a lattice
\beq
\label{fqqren}
F_{Q \bar Q,8}^{1}(a,T) = \tilde F_{Q \bar Q,8}^{1}(a,T) + 2 T d_8 \ln Z_8 =
d_8 V_3(a),
\eeq
where $V_3(r)$ is the zero temperature potential in the fundamental
representation. We use the potential derived in \cite{Necco:2001xg}.
In the matching procedure we have used Casimir scaling of the potential
at short distances. This is seen in continuum \cite{Schroder:1998vy}
and lattice \cite{Bali:2002wf} perturbation theory.  In Figure
\ref{fig:vgl_div} we show the renormalized quark-antiquark free
energies for static sources in the adjoint ($D=8$) and fundamental
($D=3$) representations together with the Casimir scaled zero temperature
potential $V_8(r) = d_8 V_3(r)$. The data clearly validates the assumption
of short distance Casimir scaling on which the procedure rests.

Once the free energies have been renormalized at small
distances, their large distance behavior is fixed. The asymptotic
value can be used to the define the renormalized Polyakov loop 
through the cluster property
\beqa
\nonumber
L^{r}_D(T) &=& \lim_{r\to\infty} \sqrt{C_{Q\overline Q,8}^{1}(r,T)}\\
   &=& \lim_{r\to\infty} \exp\left( - \frac{F_{Q\bar Q,D}^{1}(r,T)}{2
    T}\right).
\eeqa
This completes the $Q\overline Q$ renormalization procedure for the octet
loop.  The requirement that $Z_8$ depend only on $g^2$ is not explicitly
imposed in the $Q\overline Q$ renormalization procedure. However,
the results, plotted in Figure \ref{fig:renconst38} show that this is
obtained. The figure also shows that Casimir scaling of the renormalized
Polyakov loop is obtained, since $Z_8$ and $Z_3$ agree.

Any $3\times3$ unitary matrix with unit determinant is uniquely
specified by eight real numbers, which are coordinates in the abstract
group SU(3). Given these coordinates, there are canonical techniques
for building matrices in any representation $D$ which generalize
eq.\ (\ref{adjoint_link}). Hence, given the
thermal Wilson line in the fundamental, one can construct the equivalent
for arbitrary $D$. From that one can generalize every step of the
procedure from eq.\ (\ref{c1qq}) on for any $D$.


\section{Renormalization schemes}
\label{sec:scale}

\begin{figure}[thb]
\scalebox{0.65}{\includegraphics{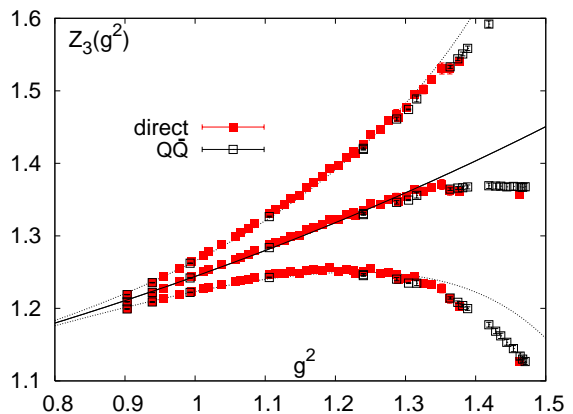}}
\caption{
Renormalization constants for fundamental loops at different scales $C=0.0,
-\sqrt{\sigma}$ (upper data points) and $\sqrt{\sigma}$ (lower data points) in
(\ref{eq_scales})
from the direct and $Q \bar Q$-renormalization
method. The solid lines are the properly scaled fit from
fig.~\ref{fig:renconst48}.
}
\label{fig_scalez}
\end{figure}

In the direct renormalization procedure the freedom of scheme choice is
the multiplicative ambiguity $L^{r}_D(T_{ref})\to
K_D L^{r}_D(T_{ref})$, for some constant $K_D$. This implies that at
another temperature the renormalized Polyakov loop is scaled by the factor
$K_D^{T_{ref}/T}$. In the $Q\overline Q$ renormalization
procedure it is the freedom of defining the zero of the $T=0$ potential
\beq
V_3(r) \longrightarrow V_3(r) + C.
\label{eq_scales}
\eeq
Using Casimir scaling for the short distance potential, this clearly leads
to the scaling
\beqa
\nonumber
L^{r}_D(T) &\to& e^{-d_DC/2T} L^{r}_D(T)\\
Z_D(g^2) &\to& e^{-d_Da(g^2)C/2},
\label{Ldscale}
\eeqa
which incorporates Casimir scaling for the Polyakov loop.
Our standard scheme choice corresponds to choosing $C$ such that the triplet
potential at $T=0$ is given by the results of \cite{Necco:2001xg}.

In Figure \ref{fig_scalez} we show examples of the
change in renormalization scheme using the two
procedures. Despite the scaling
freedom, the dependence of
$Z(g^2)$ on the bare coupling is independent of this scale and the temperature
dependence of the renormalized Polyakov loops follow from (\ref{Ldscale}).


\section{Color average free energies}\label{sec_average_cor}

\begin{figure}[tbh]
\begin{center}
\scalebox{0.65}{\includegraphics{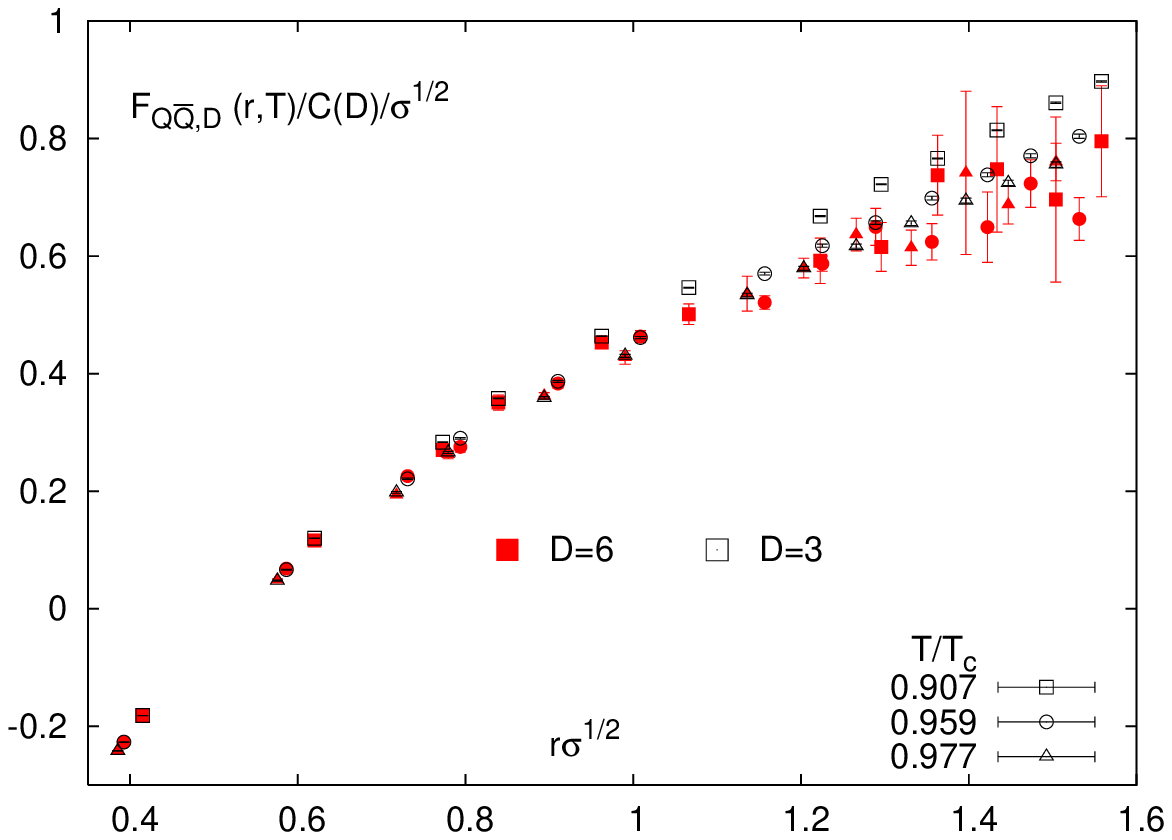}}
\scalebox{0.65}{\includegraphics{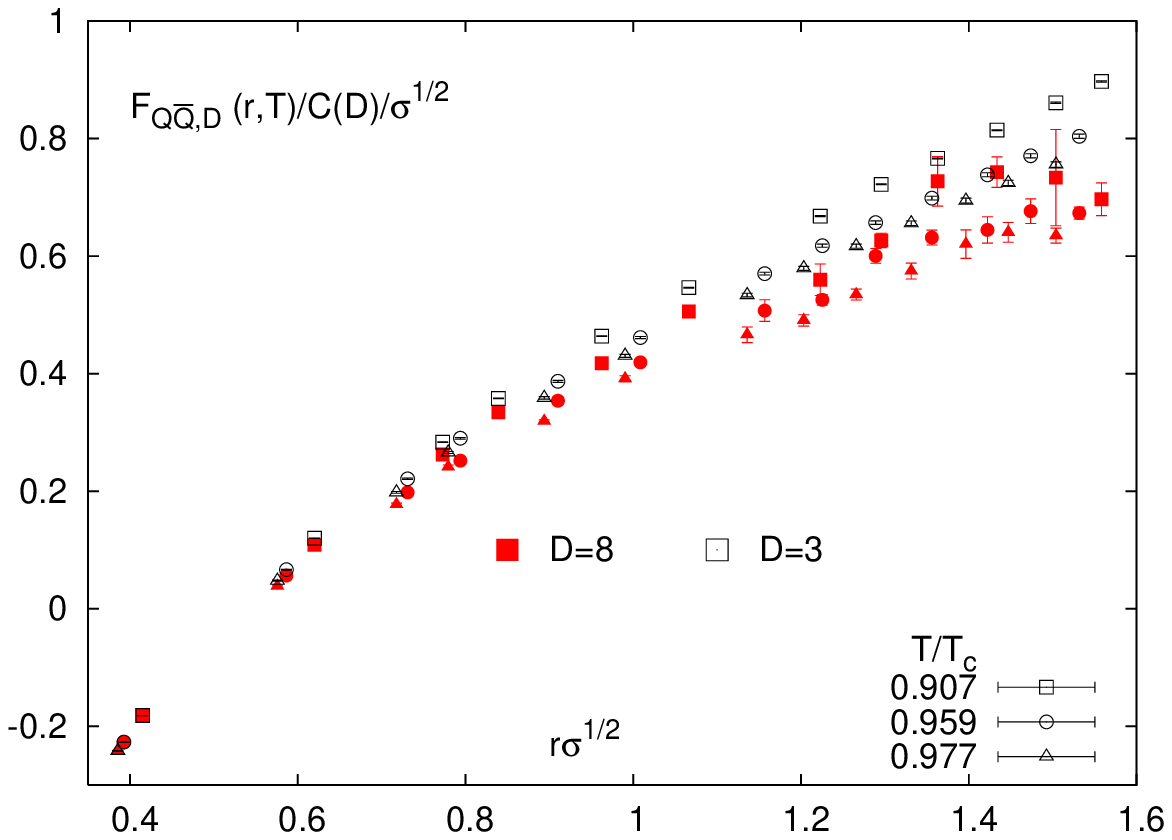}}
\end{center}
\caption{
     Average free energies for $D=6$ (top) and $D=8$ (bottom) compared to the
   fundamental free energy below $T_c$.
}
\label{fig:av_adj_sext_vgl_u}
\end{figure}

\begin{figure}[tbh]
\begin{center}
\scalebox{0.65}{\includegraphics{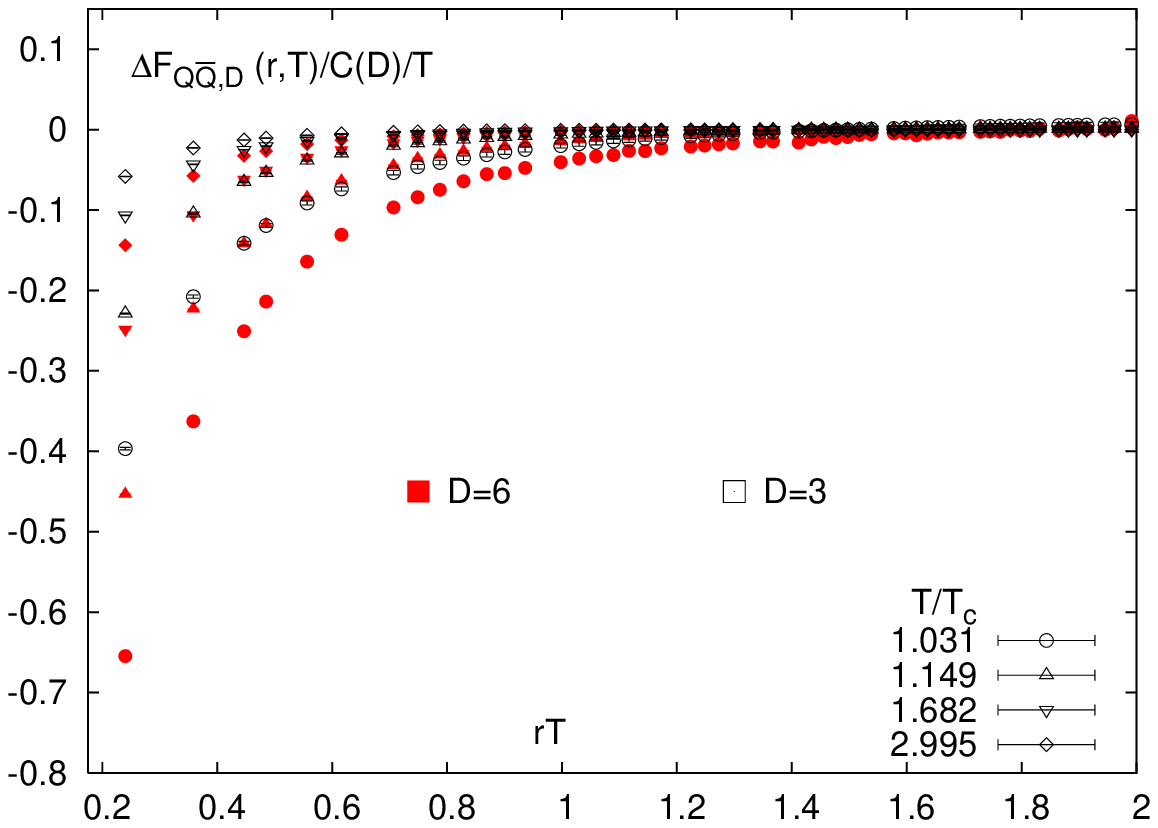}}
\scalebox{0.65}{\includegraphics{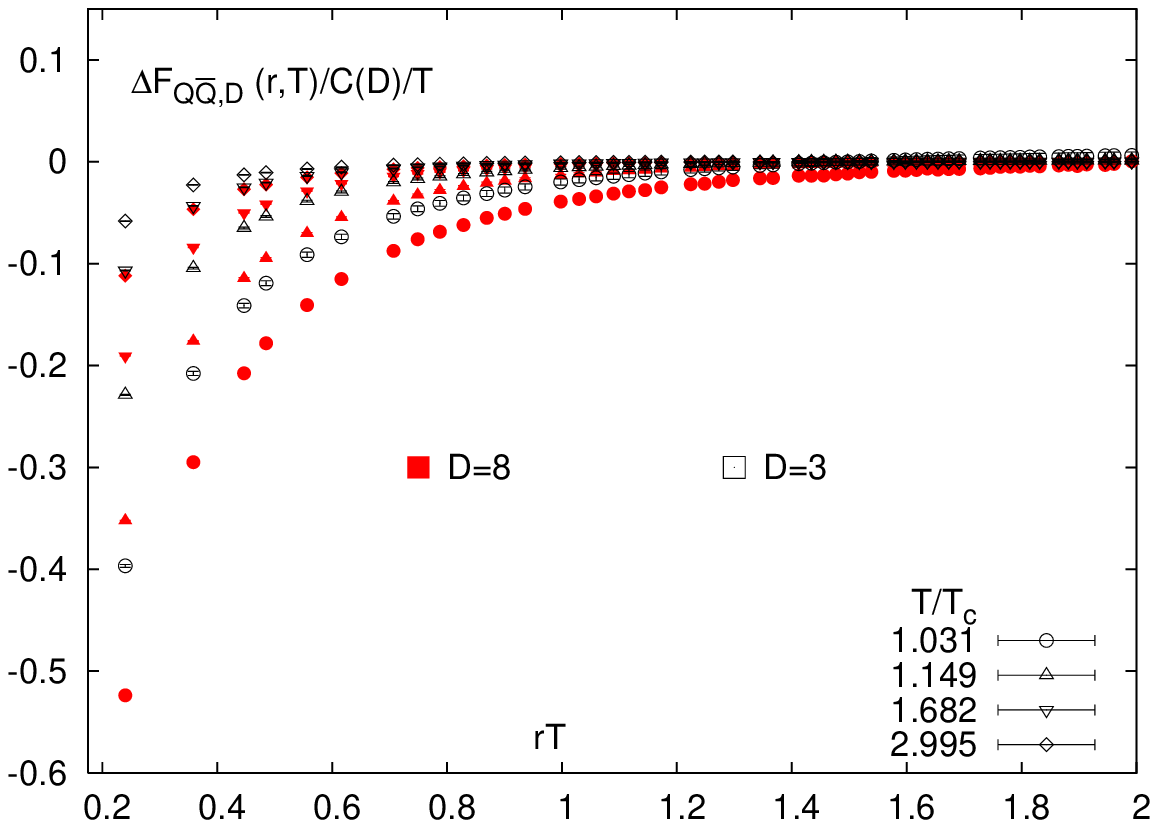}}
\end{center}
\caption{
     Average free energies for $D=6$ (top) and $D=8$ (bottom) compared to the
   fundamental free energy above $T_c$.
}
\label{fig:av_adj_sext_vgl_o}
\end{figure}

We now turn to color averaged $Q\bar Q$-free energies. Since the corresponding
correlators can be obtained without a costly gauge fixing, we were able to calculate
$F_{Q\bar Q,D}(r,T)$ for representations $D=3,6,8$ in the temperature range
$0.9-3\,T_c$ on $32^3\times 4$ lattices.
The color average correlator for at temperature $T$ in representation $D$ 
is defined by 
\begin{equation}
\label{eq:average_defa}
C_{Q\bar Q,D}(r,T)=\left\langle L^{r}_D(x_1)L^{r*}_D(x_2))\right\rangle,
\end{equation}
where the star denotes complex conjugation.
The color average free energy $F_{Q\bar Q,D}(r,T)=-T\log C_{Q\bar Q,D}(r,T)$.
If Casimir scaling 
holds, then
\begin{equation}
\label{eq:average_cs}
F_{Q\bar Q,D}(r,T)/C_2(D) =
F_{Q\bar Q,D'}(r,T)/C_2(D').
\end{equation}
We test this relation below and above $T_c$.

Below $T_c$ we employ the renormalized average
free energies 
\begin{equation}
\label{eq:average_defb}
F_{Q\bar Q,D}(r,T)=\tilde F_{Q\bar Q,D}(r,T) + 2 T d_D \ln Z_D(g^2)
\end{equation}
for the three lowest temperatures divided by their Casimir
in fig.~\ref{fig:av_adj_sext_vgl_u}.
The renormalization constants used are those
found by the renormalization procedure described in sec.~\ref{sec_fundamental}.
Thus for the smallest distances all curves 
coincide as a consequence of the renormalization
procedure.
However, for all $T<T_c$ and representations $D=6,8$ deviations to smaller
values start to show up quite early, i.~e.~
for separations $r\sqrt{\sigma}\gsim 1$ for $D=6$
and $r\sqrt{\sigma}\gsim 0.8$ for $D=8$, respectively.
This effect is more pronounced for the adjoint average free energy than for the sextet.
The effect of string breaking sets in at larger distances than shown here and will
be discussed in sec.~\ref{sec_strinbreaking}.

Above $T_c$ we compare 
\begin{equation}
\label{eq:average_defc}
\Delta F_{Q\bar Q,D}(r,T)=F_{Q\bar Q,D}(r,T)-F_{Q\bar Q,D}(r\to\infty,T)
\end{equation}
divided by their Casimir for the
same representations $D=3,6,8$ in fig.~\ref{fig:av_adj_sext_vgl_o}.
We observe {\em screening} to take place in both higher representations.
The curves for both $D=6$ and $D=8$ deviate
to smaller values
compared to the fundamental case. We find that the ordering
\begin{equation}
\label{eq:average_ordering}
\frac{\Delta F_{Q\bar Q,6}(r,T)}{C_2(6)}<\frac{\Delta F_{Q\bar Q,8}(r,T)}{C_2(8)} <
\frac{\Delta F_{Q\bar Q,3}(r,T)}{C_2(3)}<0,
\end{equation}
holds throughout the entire distance interval above $T_c$.

Thus, we conclude, that Casimir scaling \eqref{eq:average_cs} is clearly violated 
for the average $Q\bar Q$ free
energies in the temperature range $0.9-3\,T_c$ for the fundamental, sextet and
adjoint representations.


\end{document}